# Exploring the catastrophic regime: thermodynamics and disintegration in head-on planetary collisions

Jingyao Dou[1]★, Philip J. Carter[1], Simon Lock[2] and Zoë M. Leinhardt[1]

[1]*School of Physics, H.H. Wills Physics Laboratory, University of Bristol, Bristol BS8 1TL, UK*
[2]*School of Earth Sciences, University of Bristol, Bristol BS8 1RJ, UK*

## ABSTRACT

Head-on giant impacts (collisions between planet-sized bodies) are frequently used to study the planet formation process as they present an extreme configuration where the two colliding bodies are greatly disturbed. With limited computing resources, focusing on these extreme impacts eases the burden of exploring a large parameter space. Results from head-on impacts are often then extended to study oblique impacts with angle corrections or used as initial conditions for other calculations, for example, the evolution of ejected debris. In this study, we conduct a detailed investigation of the thermodynamic and energy budget evolution of high-energy head-on giant impacts, entering the catastrophic impacts regime, for target masses between 0.001 and 12 $M_\oplus$. We demonstrate the complex interplay of gravitational forces, shock dynamics, and thermodynamic processing in head-on impacts at high energy. Our study illustrates that frequent interactions of core material with the liquid side of the vapour curve could have cumulative effects on the post-collision remnants, leading to fragmentary disintegration occurring at lower impact energy. This results in the mass of the largest remnant diverging significantly from previously developed scaling laws. These findings suggest two key considerations: (1) head-on planetary collisions for different target masses do not behave similarly, so caution is needed when applying scaling laws across a broad parameter space; and (2) an accurate model of the liquid-vapour phase boundary is essential for modelling giant impacts. Our findings highlight the need for careful consideration of impact configurations in planetary formation studies, as head-on impacts involve a complex interplay between thermodynamic processing, shocks, gravitational forces, and other factors.

## 1 INTRODUCTION

Giant impacts, collisions between planet-sized bodies, are common in the late stage of planet formation. Previous studies have shown that giant impact(s) can successfully explain key properties of the Moon (Canup & Asphaug 2001; Canup 2004; Ćuk & Stewart 2012; Nakajima & Stevenson 2014; Lock et al. 2018; Ruiz-Bonilla et al. 2020; Kegerreis et al. 2022; Timpe et al. 2023; Yuan et al. 2023), Uranus (Kegerreis et al. 2018; Reinhardt et al. 2020), Saturn (Teodoro et al. 2023), Mercury (Benz et al. 2007; Chau et al. 2018; Reinhardt et al. 2022) and several known exoplanets (Bonomo et al. 2019; Kenworthy et al. 2023; Naponiello et al. 2023). The outcomes of giant impacts are complicated and are commonly studied using smoothed particle hydrodynamics (SPH) simulations. Previous works using SPH simulations have studied the collisional stripping outcome between different-sized bodies (Marcus et al. 2009; Leinhardt & Stewart 2012; Movshovitz et al. 2016; Carter et al. 2018; Reinhardt et al. 2022; Dou, Carter & Leinhardt 2024), the effect of SPH resolution and equation of state (EoS) (Kegerreis et al. 2018; Meier, Reinhardt & Stadel 2021), the erosive loss of planetary atmospheres (Denman et al. 2020; Kegerreis et al. 2020; Denman, Leinhardt & Carter 2022; Lock & Stewart 2024), extreme debris discs (Watt, Leinhardt & Su 2021; Lewis, Watt & Leinhardt 2023; Watt, Leinhardt & Carter 2024), and built data sets including various impact configurations (Cambioni et al. 2019; Gabriel et al. 2020; Winter et al. 2023; Emsenhuber et al. 2024).

There are numerous factors that can affect the outcome of a giant impact, and the range of parameters involved is inherently large. Important variables include the mass of the target, the ratio of impactor to target mass ($\gamma$), the speed of the impact, the angle at which the impact occurs, the spin rate (or alternatively angular momentum) of the colliding bodies. Considering how computationally expensive SPH simulations are, it would be unfeasible to explore all of these parameters in a single study.

To simplify the analysis and save computing resources, many previous studies have often utilized the head-on impact setup as a simplification or end-member scenario. Recent studies have shown that high-energy head-on impacts can produce diverse outcomes. Reinhardt et al. (2022) proposed new scaling laws based on simulations of head-on impacts between super-Earths at high enough impact velocities that the impacts form very dense remnants. They showed that in the catastrophic disruption regime, above a certain normalized impact energy, the mass of the largest remnant decreases rapidly and the iron mass fraction increases rapidly, both following a power law as a function of normalized impact energy. These new scaling laws deviate significantly from those proposed by Marcus et al. (2009) and Carter et al. (2018) at high-impact energies. The cause of this trend of rapidly decreasing remnant mass remains unclear. Dou et al. (2024) reported that in head-on giant impacts with equal-mass objects ($\gamma = 1.0$), fragmentary disintegration occurred at lower normalized impact energy as the impact velocity increased (refer to fig. 5 in their paper). This fragmentary disintegration leads to a sharp change in the mass of the largest remnant, as observed by Reinhardt et al. (2022). An impact is defined to be in the super-catastrophic regime when the mass of the largest remnant normalized to the total colliding mass, $M_{\rm lr}/M_{\rm tot}$, is less than 10 per cent (where $M_{\rm lr}$ is the mass of the largest remnant and $M_{\rm tot}$ is the initial total mass of the system). Leinhardt & Stewart (2012) found and proposed that impacts tend to be in the super-catastrophic regime when normalized impact energy $Q_{\rm R}/Q_{\rm RD}^*$ ($Q_{\rm R}$ is impact energy and $Q_{\rm RD}^*$ is the impact energy needed to disperse half of the system's total mass) is larger than 1.8. Above this energy, the mass of the largest remnant decreases more slowly but smoothly with increasing impact energy. However, Dou et al. (2024) demonstrated that an onset of fragmentary disintegration can occur when $Q_{\rm R}/Q_{\rm RD}^*$ is as small as 1.2 and $M_{\rm lr}/M_{\rm tot}$ is as large as 40 per cent, indicating that disintegration and the transition to super-catastrophic disruption can happen at lower energies than previously thought for head-on impacts.

★ E-mail: qb20321@bristol.ac.uk

The head-on impact configuration represents the most extreme case, where the target and impactor are perfectly aligned, and material from both bodies are equally actively involved in the impact. However, it is important to note that head-on impacts are actually the rarest configuration in nature. According to the probability of impact angle ($P \propto \sin(2\theta_{\rm imp})$, Shoemaker 1962), oblique impacts are far more common. Furthermore, Dou et al. (2024) demonstrated that kinetic momentum transfer and vapourization-induced ejection behave differently in head-on versus oblique impacts. Specifically, the dynamics of material ejection and the subsequent thermal and mechanical evolution of the impactor and target differ significantly between the head-on and oblique scenarios. Consequently, while head-on impacts provide valuable insights into impact processes, it is important to bear in mind that the results derived from such studies might not be directly applicable to oblique impacts.

Although head-on impacts are rare and 'idealized', the extreme impact conditions they involve provide an opportunity to study numerical stability, consistency, and the thermodynamic consequences of giant impacts. In this work, to better understand the fragmentary disintegration behaviour reported by Dou et al. (2024), we focus on tracking and comparing the thermal properties and energy budgets of giant impacts across a range of impact energies. This approach has allowed us to analyse how certain thermodynamic processes alter the final outcome of giant impacts and how energy is exchanged during the collision. In sum, this work demonstrates how head-on impacts behave differently at varying impact energies, suggesting that conclusions derived from head-on impacts should be treated with caution when applying to oblique impacts. When fitting or training scaling laws to predict the mass and iron mass fraction of the remnants after a giant impact for using in *N*-body simulations, it is advisable to treat data from head-on impacts separately to achieve a more robust prediction model.

This paper is organized as follows. In Section 2, we describe the methods used to create planetary bodies and conduct giant impact SPH simulations. We also explain the tools and methods used to analyse the simulation results. In Section 3, we present the results of head-on impacts with target masses ranging from 0.001 to approximately 12 $M_\oplus$. We then demonstrate how thermodynamic processes and energy exchange behave differently in head-on impacts at 1.0 $M_\oplus$ across a large spectrum of impact velocities, covering the entire range of impact energies. In Section 4, we discuss the results of nearly head-on and oblique impacts. We also examine the trigger conditions for fragmentary disintegration and the influence of the density floor (see definition of density floor in Section 2.2) in SPH simulations on the results. Finally, in Section 5, we summarize the key findings of the paper.

## 2 METHODS

We focused on head-on impacts with target masses ($M_{\rm targ}$) ranging from 0.001 to approximately 12 $M_\oplus$. We investigated high-energy impacts where material strength plays a minimal role, and did not consider material strength even for the smallest target mass of 0.001 $M_\oplus$ (around 720 km in radius). Our primary analysis utilizes simulation results from equal-mass impacts, as these scenarios exhibit the most extreme interactions of shock and gravitational forces, providing a clearer understanding of the factors contributing to the impact outcomes.

SPH simulations were performed using SWIFT (v 0.9.0, branch: planetary_plus_subtask_speedup,[1] Kegerreis et al. 2019; Schaller et al. 2023) with a 'vanilla' form of SPH (Price 2012) and the Balsara (1995) switch for the artificial viscosity. We used a 3D cubic spline kernel with 48 nearest neighbours, corresponding to a ratio of smoothing length to inter-particle separation of 1.2348 (Dehnen & Aly 2012). We investigated the influence of the choice of SPH kernel on our results by testing the Wendland $C^2$ and Wendland $C^6$ kernels with 100 and 400 nearest neighbours, respectively. The results of these tests indicated that the selection of the SPH kernel did not have a significant effect on the outcomes of our simulations. The default artificial viscosity parameters for the Monaghan (1992) model are set to $\alpha = 1.5$ and $\beta = 2\alpha$ (Reinhardt & Stadel 2017). We used a Courant factor of 0.2.

### 2.1 Initial conditions

The initial condition setup follows Dou et al. (2024): two-layered planets are generated using WOMA (Kegerreis et al. 2019; Ruiz-Bonilla et al. 2020) with isentropic temperature profiles. The initial planets were all differentiated with 30 per cent iron core and 70 per cent forsterite mantle. We use the iron (Stewart et al. 2019; Stewart 2020) and forsterite (Stewart 2020) M-ANEOS equations of state, but re-generated the EoS tables (Dou 2024) with higher maximum densities (100 g cm$^{-3}$ for forsterite and 200 g cm$^{-3}$ for iron). For target masses above 0.1 $M_\oplus$, we select entropies of core and mantle such that each layer is in solid phase while close to the melt curve. The phase boundaries and critical points we use in the results section are all from Stewart et al. (2019) and Stewart (2020).

Lots of head-on impacts with target masses of approximately one Earth mass are studied in this work. Therefore, here we provide detailed profile information for the target planet: the mass is 0.999 $M_\oplus$ and the radius is 1.03 $R_\oplus$. The mantle entropy is set to 3027 J K$^{-1}$ kg$^{-1}$ and the core entropy is set to 1750 J K$^{-1}$ kg$^{-1}$. The surface temperature is approximately 2165 K. Fig. 1 shows the thermal profile of the planet. To assess the impact of the initial thermal profile on our results, we conducted a subset of simulations with a surface temperature of 300 K and a core-to-mantle temperature difference within 200 K. The results of these tests indicated that the choice of the initial thermal profile did not significantly influence the outcomes of our simulations.

---
[1] https://github.com/SWIFTSIM/SWIFT/tree/planetary_plus_subtask_speedup

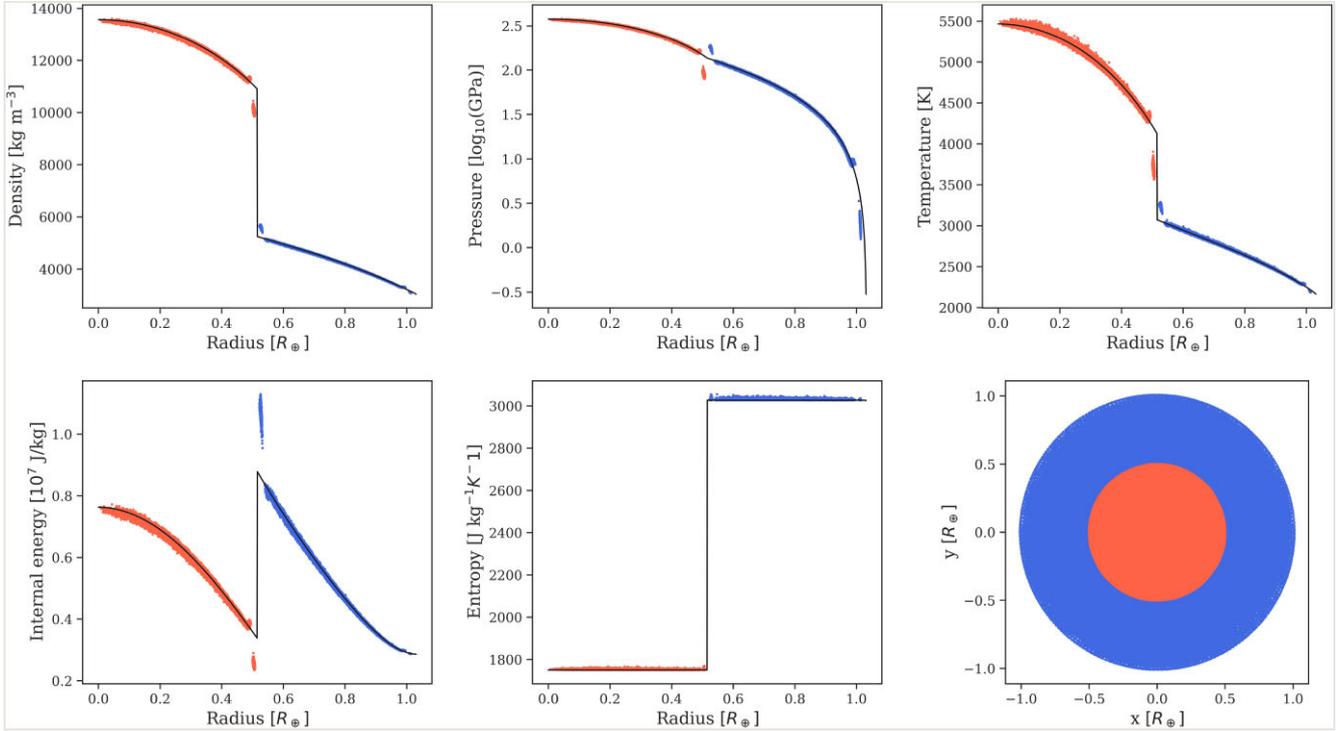

**Figure 1.** Particles radial profile for the target planet with a mass of approximately 1 $M_\oplus$ after equilibration process. The red points represents the iron core and the blue represents the forsterite mantle. Solid black line shows the analytical radial profile calculated from WoMa. The right-most panel in the second row shows a slice of the equilibrated planet in the middle plane ($z = 0$).

For target planets with masses of 0.001 and 0.01 $M_\oplus$, we model the core and mantle as fully molten with entropies of 2000 and 3550 J K$^{-1}$ kg$^{-1}$, respectively. We chose to use fully molten planets as, in this small target mass regime, impact shocks are very low in magnitude. If solid planets were used, even high-velocity impacts would only partially melt the colliding bodies. However, as material are released from shock, the treatment of material as fluids in SPH calculations forces the particles to become less dense. The only way they can achieve this is by reaching very low pressure. While the material remains solid there is only a slight expansion due to the limited thermal expansion of solids. Consequently, SPH particles will exhibit unrealistically low pressure until they sublimate. We conducted a test group of simulations using solid target planets at 0.001 and 0.01 $M_\oplus$ and found that the thermal state has a trivial effect on the mass and iron mass fraction of the largest remnant. However, it certainly affects the thermodynamic history of planetary bodies.

For most analyses, the particle resolution in the one Earth mass target planet is 498 024 particles, resulting in a total resolution of approximately one million particles for equal-mass impact simulations. The particle resolution for other target mass planets used in this work is around 200 000 particles. We discuss the influence of SPH resolution on our results in Section 4.5.

Before the impact simulations, each body was equilibrated in a cooling simulation for 20 h of simulation time in isolation to reach a stable hydrostatic state. During the first 10 h, the entropy of the core and mantle were fixed to the desired values at each time step

(see Carter et al. 2018) in order to produce planets with isentropic material layers. Over the next 10 h, the planets evolved without this additional damping, towards a hydrostatic profile. After the equilibration process, particles have a root-mean-squared velocity that is less than 1 per cent of the planet's escape velocity.

For equal-mass head-on impacts, we considered impact velocities ranging from 1.0 to around 3.0 times the mutual escape velocity ($V_{esc}$). For impacts with impactor-to-target mass ratio, $\gamma$, less than one, the impact velocity is increased until the summed mass of resolved post-impact remnants is just below 10 per cent of the total mass of the initial system in the fastest impact simulation. The mutual escape velocity is defined by

$$V_{esc} = \sqrt{\frac{2G(M_{targ} + M_{imp})}{R_{targ} + R_{imp}}}, \quad (1)$$

where $G$ is the gravitational constant, and $M_{targ}$, $R_{targ}$ and $M_{imp}$, $R_{imp}$ are the mass and radius of the target and the impactor, respectively.

We set up each simulation with target and impactor planets initially separated by a distance such that contact occurs one hour after the start of the simulation to allow tidal deformation of the SPH planets (Kegerreis et al. 2020). For most head-on impact simulations, the simulation time was approximately 20 h, during which the mass and iron mass fraction of the largest remnant exhibited little change after 10 h. The simulations were run in cubic boxes with side lengths of 1000 $R_\oplus$ or larger for high-velocity impacts. Any particles leaving the box were removed from the simulation.

## 2.2 SPH simulation density floor

The density, $\rho$, of an SPH particle is estimated as a weighted sum of the masses of the neighbouring particles:

$$\rho(\mathbf{r}_i) = \sum_j^{N_{ngb}} m_j W(\mathbf{r}_i - \mathbf{r}_j, h_i), \quad (2)$$

where $m_j$ is the neighbour particle mass, $W(\mathbf{r}, h)$ is the smoothing kernel and $h$ is the smoothing length. SWIFT uses a parameter $h_{max}$ (see SWIFT documentation[2]) setting the maximal allowed smoothing length of SPH particles. Particles in the simulations can only have smoothing length less than or equal to $h_{max}$, which in turn sets a density floor (minimum density of SPH particles). In a three-dimensional SPH calculation, $W(\mathbf{r}, h)$ can be expressed as

$$W(\mathbf{r}, h) = H^{-3} w(|\mathbf{r}|/H) = H^{-3} C \left( \left(1 - \frac{|\mathbf{r}|}{H}\right)^3 - 4\left(\frac{1}{2} - \frac{|\mathbf{r}|}{H}\right)^3 \right), \quad (3)$$

where $H$ is the kernel-support radius, and according to Dehnen & Aly (2012), for the 3D cubic spline kernel we used in our simulations, $C = 16/\pi$ is a constant, and $H/h = 1.825\,742$. Therefore, for the case where a particle has zero nearest neighbours included in the smoothing kernel, the lowest density an SPH simulation can resolve is then dependent on the maximum smoothing length and particle mass as

$$\rho_{floor} = m_i (1.825742\, h_{max})^{-3} \times \frac{16}{\pi} \times \left(1 - 4 \times \left(\frac{1}{2}\right)^3\right)$$
$$\approx 0.41843\, m_i h_{max}^{-3}. \quad (4)$$

In SPH simulations, all the particles have roughly the same mass, hence, the density floor is affected by both $h_{max}$ and the number of particles in the simulation.

The use of a finite $h_{max}$ prevents particles from interacting with each other at very large separations and keeps particles from reaching extremely low densities in poorly resolved regions. The advantage of a finite $h_{max}$ is that it optimizes parallel computing by reducing the need for computationally expensive searches for distant neighbouring particles. A maximum smoothing length, or a minimum SPH density, can have a substantial influence on the post-collision disc (Hull et al. 2023) as relatively few particles are left in the disc region. Most of the simulations in this study used an $h_{max} = 0.2\,R_\oplus$ resulting in a density floor of $\sim 0.0024\,g\,cm^{-3}$ for simulations with $1.0\,M_\oplus$ and 498 024 particles in the target. For other target mass impacts, we keep the density floor to be around $\sim 0.0037\,g\,cm^{-3}$ by varying $h_{max}$ accordingly. For a given particle resolution, the larger the $h_{max}$ the longer a simulation will take to run. Since this study focuses on the large-scale properties of post-collision remnants (rather than the properties of the disc), our smaller value for $h_{max}$ provides an acceptable balance between computing cost and accuracy. We discuss the influence of the density floor on our results in Section 4.6.

## 2.3 Simulation analysis

### 2.3.1 Search for bound particles

We utilized the same remnants searching algorithm as in Dou (2023), as described in Marcus et al. (2009) and Carter et al. (2018), to identify particles bound to a post-collision remnant. Initially, the potential and kinetic energies of the post-collision particles were calculated with respect to the seed particle that was closest to the potential minimum. Then, the gravitationally bound particles were selected, and the centre-of-mass position and velocity of all the bound particles were recomputed and used as the seed for the next iteration. This process was repeated for the remaining unbound particles until convergence was achieved, i.e. no particles were included in or removed from the largest remnant between subsequent iterations. We did not consider the re-accretion of ejecta that were unbound to the largest post-collision body but bound to the star, similar to previous studies (Marcus et al. 2009; Leinhardt & Stewart 2012; Carter et al. 2018; Reinhardt et al. 2022) as this is highly sensitive to the architecture of the host system. The iron mass fraction of a remnant is the ratio of the mass of bound iron particles to the remnant's total mass.

### 2.3.2 Thermodynamic tracking

Due to the high frequency of output snapshots, that record the position and properties of each SPH particle, and the high numerical resolution of particles used in the simulations, tracking the thermodynamic properties of all SPH particles over time would be computationally unfeasible and require too much memory. Instead, we select and monitor a subset of particles in the target planet only. In equal-mass head-on impacts, the target[3] and impactor material undergo similar shock and thermodynamic processes; therefore, core particles from the target planet should be representative of all core particles.

We divide the core and mantle into 40 shells based on the particles' radius and randomly select 20 per cent of particles in each shell to track. For impacts with the same target mass, the tracked particles are the same for simulations with different impact velocities, facilitating direct comparison. This method ensures that the tracked particles' position represents the entire target's particles well, and the shock and thermodynamic processes experienced by particles in different regions are accurately tracked. To assess the sensitivity of our results to the number of particles used in each shell, we re-ran a subset of the analysis using 50 per cent of the particles in each shell. The results obtained from these tests did not show significant differences compared to our original findings.

### 2.3.3 Energy tracking

Following Carter, Lock & Stewart (2020), we define the giant impact energy budget at any moment in time by three terms: kinetic energy, internal energy, and a 'participating' potential energy term. This latter term is determined by subtracting the minimum value of the potential energy at any time in the simulation from the current potential energy: $E_{pot} - E_{pot,min}$. The potential energy usually reaches its minimum between 1 and 1.5 h after the start of the simulations, as the colliding bodies are initially separated to ensure the collision occurs at 1 h. This energy is more negative than the potential energy at either the beginning or end states. The offset factor $E_{pot,min}$, converts the participating potential energy term to a positive value. The total participating energy of an impact is the sum of kinetic energy, internal energy, and participating potential energy.

### 2.3.4 Energy analysis

The catastrophic disruption criteria, $Q_{RD}^*$, is the impact energy needed to permanently disperse half of the system's total mass. For our head-on impacts, we determine the critical specific impact energy for catastrophic disruption by linear interpolation between the two data points that bound the specific impact energy where half of the total colliding mass remains in the largest remnant. The specific reduced impact energy $Q_R$ is defined by

$$Q_R = 0.5 \mu \frac{V_i^2}{M_{tot}}, \tag{5}$$

where $M_{tot} = M_{targ} + M_{imp}$ is the total mass of the system, $\mu = M_{targ} M_{imp}/M_{tot}$ is the reduced mass, and $V_i$ is the impact velocity. Therefore, $M_{lr}/M_{tot} = 0.5$ (where $M_{lr}$ is the mass of the largest post-collision remnant) for a collision with $Q_R = Q_{RD}^*$.

---

[2] https://swift.strw.leidenuniv.nl/docs/index.html

[3] Before the collision, the centre of mass (CoM) of the system is placed at the origin, with the target planet located on the negative $x$-axis.

The gravitational binding energy $E_{grav.bind}$ of the system is another significant factor influencing the results of a giant impact. Rather than using an analytical equation to determine the binding energy of targets and impactors, we directly use the potential energy from our equilibrated SPH planets. We calculate the system's total gravitational binding energy as follows:

$$E_{grav.bind} = -\frac{G M_{targ} M_{imp}}{R_{targ} + R_{imp}} + E_{bind,targ} + E_{bind,imp}. \tag{6}$$

### 2.4 Catastrophic and super-catastrophic impacts

In this work, we define super-catastrophic impacts as those in which the post-collision structure undergoes fragmentary disintegration as well as impacts which result in no dominant remnant such that only small debris pieces are left in the system, each having a mass less than 10 per cent of the total system mass. Catastrophic impacts, on the other hand, are characterized by impacts that produce a largest remnant with mass ($M_{lr}/M_{tot}$) of 0.5 or less, occurring before the onset of super-catastrophic impacts.

### 2.5 Impedance-match calculations

To calculate the maximum entropy increase that could be achieved by a single shock from the initial impact, we used semi-analytic impedance match calculations. To provide a simple upper estimate, we assumed a planar shot from the surface to the centre of the planet. The initial shock pressure at the surface was calculated by the impedance match between the mantle of the colliding bodies at the impact velocity. The shock was then propagated down through the mantle to the core–mantle boundary (CMB) by treating the pressure gradient in the mantle as a series of 100 pressure/density steps with impedance match performed at the boundary of each. After calculating the impedance match between the core and mantle at the CMB, the shock was then propagated to the centre of the planet in the same manner with 100 steps. Changing the number of steps had little effect on the final result. The same EOS were used as for the SPH simulations. We used surface pressure, CMB pressure and centre core pressure from the profile generated by WoMa to calculate impedance match results of impacts at various impact velocities for the $1.0\,M_\oplus$ target.

## 3 RESULTS

In this section, we demonstrate how planets of varying masses behave differently upon entering the catastrophic impact regime for head-on impacts. We present the distinct thermodynamic and shock histories experienced by core particles. Specifically, we focus on studying and tracking the evolution of core material, as high-energy impacts lead to significant vaporization of the cores of the colliding bodies, resulting in greater thermodynamic variability of core material compared to mantle material. The influence of mantle material is further discussed in Section 4.2. Additionally, we conduct a detailed analysis of head-on, equal-mass impacts of $1.0\,M_\oplus$ bodies to enhance our understanding of the underlying causes of fragmentary disintegration (rapid change of the mass of the largest remnant with increasing of impact energy above a threshold).

### 3.1 Fragmentary disintegration

Fig. 2 displays the bound particles of post-collision remnants in the $y$–$z$ plane for impacts with various target masses (columns) at velocities just below (top panels) and just above (bottom panels) the velocity needed to trigger fragmentary disintegration. In the top row, fragmentary disintegration has not yet occurred, leaving one remnant (represented by a single colour) that contains most of the bound mass. In the bottom row, fragmentary disintegration has occurred, resulting in the formation of multiple similarly massed remnants (represented by multiple colours). The largest remnants of impacts at target masses of 0.1 and $1.0\,M_\oplus$, at velocities below that required for fragmentary disintegration, display a ring structure (as seen in the middle two sub-panels of the top row in Fig. 2). Given more simulation time, the particles forming the ring will collapse towards the centre, creating a puffy cylinder structure with a dense central region and forming a single large remnant, similar to the structure shown for a target mass of $0.01\,M_\oplus$ (the leftmost sub-panel of the top row). However, a slight increase in impact velocity (second row in Fig. 2) tends to destabilize the ring structure, causing it to disintegrate into several smaller remnants. The onset of fragmentary disintegration is sensitive to the impact velocity for all listed target masses, and the mass ratio of the largest remnant just before fragmentary disintegration also varies at different target mass impacts.

At higher target masses, as demonstrated by impacts of around $11.86\,M_\oplus$ (rightmost sub-panels in the top and bottom row of Fig. 2), the central region exhibits a web-like structure with dense strings and sparse voids in between. At velocities below fragmentary disintegration, the largest remnant is situated in the most central region of this web. The fragmentary disintegration process in these cases involves the breaking up of this web-like structure rather than the ring structure observed in lower mass targets.

For target masses at $0.01\,M_\oplus$, fragmentary disintegration of the largest remnant begins when $M_{lr}/M_{tot}$ is around 10 per cent, which is close to the threshold for the previously defined super-catastrophic regime. However, when the target mass is between 0.01 and $1.0\,M_\oplus$, a mass range frequently studied previously, fragmentary disintegration can occur when the normalized impact energy $Q_R/Q_{RD}^*$ is as small as 1.2 and the largest remnant mass is just above 40 per cent. As the target mass increases to $11.86\,M_\oplus$, the largest remnant mass right before disintegration drops back to around 15 per cent, and the bound remnants show irregular filamentary shapes, making it difficult to determine whether we are observing the same fragmentary disintegration processing as shown for lower target masses. For masses below one Earth mass, the breakup of the central ring structure is a new regime of disruption and the mass of the largest remnant follows a different trend with increasing impact velocity for head-on impacts (Section 3.4).

In these high-mass, highest energy impact scenarios, the collision is so energetic that it disrupts the entire system, preventing material from collapsing toward the centre. Therefore, the post-collision structure of higher target masses impacts evolve with slower speed. In addition, as tested by Dou et al. (2024), *N*-body simulations suggest that these smaller fragments after the fragmentary disintegration will move away from each other and will not re-accrete or merge together again.

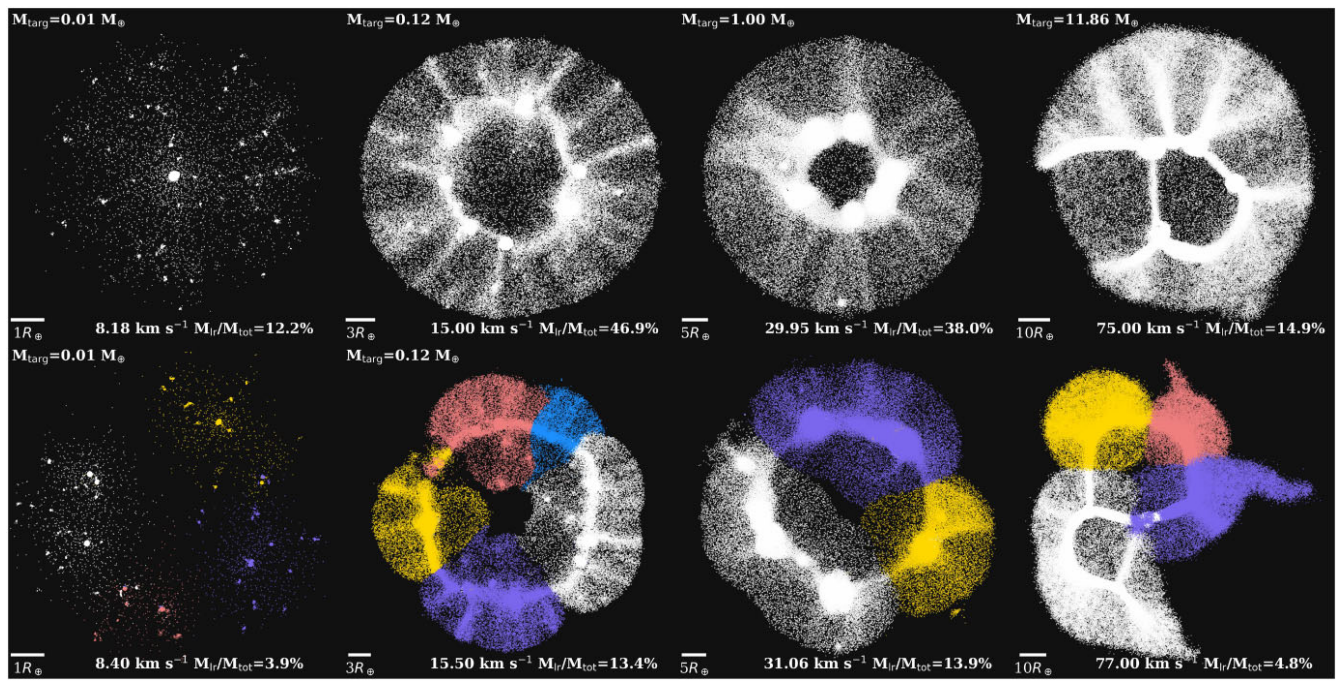

**Figure 2.** Snapshots of simulations that produced a classic largest remnant (top row) and those at slightly higher velocities that produced disintegrated remnants (bottom row) are shown for target planet masses ranging from 0.01 to 12 $M_\oplus$ in the *y*-*z* plane. White particles are SPH particles that were bound to the largest remnant in all panels. Initially, the two colliding bodies were moving along the *x*-axis (i.e. into/out of the page). All depicted impacts are equal-mass head-on impacts, simulated for 20 h. Impact velocities and the mass ratio of the largest remnant are given in the bottom right of each sub-panel. In the bottom panels, each colour represents a bound remnant (white, indigo, yellow, red, and blue particles show remnants with masses in descending order, respectively). The mutual escape velocities for the listed low to high target mass impacts are 2.254, 5.327, 11.093, and 27.759 km s$^{-1}$, respectively.

### 3.2 Spatial distribution of iron

Unlike impacts occurring at mutual escape velocity, where little core material is disturbed, high-impact velocities result in significant compression and disturbance of all material. Compared to core material, mantle material has a lower density and is therefore easier to eject. Additionally, at high-impact energies, core vapourization begins to significantly contribute to the mantle stripping process (Dou et al. 2024). Consequently, in high-velocity head-on impacts, it is dominantly the core material that define the post-collision structure and hence affects how the largest remnant evolves.

Fig. 3 shows how the spatial distribution of core material evolves over time and varies with different target masses. We selected impacts with velocities where fragmentary disintegration just begins to occur. For a target mass of 0.001 $M_\oplus$, since there is no clear evidence of fragmentary disintegration, we selected an impact velocity where the final $M_{lr}/M_{tot}$ is around 8 per cent. Notably, at slightly lower velocities, which correspond to the velocity of the top row in Fig. 2, the general structure of core material spatial distribution remains very similar without any major differences. In Fig. 3, however, from the top row to the bottom row, as the target mass increases, the post-collision structure changes significantly.

For impacts at or below one Earth mass and above 0.01 $M_\oplus$, core material tends to initially form a ring structure in the central (inner 3–5 $R_\oplus$) region and a web structure with void spaces around it over time (the third and fourth columns in Fig. 3). The void spaces become larger as they extend further from the impact point. Surrounding material then begins to accumulate on the ring, collide, and merge back to the centre. The lower the target mass, the earlier the post-collision structure starts to display a ring-like shape. The outer disc near the edge and the inner central region are separated by a gap with comparatively low density. For a target mass of 0.001 $M_\oplus$, there is no appearance of a ring structure, and the density of core material decreases gradually from the centre to the edge. For impacts at 11.86 $M_\oplus$, there is also no obvious ring structure in the central region. Instead, core material spreads out more evenly, without a gap between the edge of the disc and the centre. Compared to impacts with lower target masses, the evolution of the core material in an impact with a target mass of 11.86 $M_\oplus$ is less pronounced and shows less variation over time. As discussed in Section 3.3, for collisions with high target masses at high energy approaching the super-catastrophic regime, the initial shock is so strong that it dominates the final state of the impacts.

The core material in the snapshots with target mass above 0.01 $M_\oplus$ all exhibit web-like structures. In the first 1–2 h of the collision, numerous high-density 'bulbs' are generated, which later evolve to form the web structure. This bulb generation results in regional variations in material density. Over time, the high-density regions slowly shape into the arms or strings of the web.

During head-on impacts, when the impactor and target first make contact material near the impact site is ejected in largely vapourized ejecta plumes. Quickly, a shock wave propagates from the impact point into both the target and impactor. The breakout of the shock from the core to the mantle provides a shock-kick that ejects material from the system (an analogous process described for atmospheric loss in Lock & Stewart 2024). Due to the lower shock impedance of the forsterite mantle compared to the iron core, the pressure in the shocked mantle is lower than in the shocked core. The core must release to a lower pressure, following an isentrope (as shown in Fig. 4), until it intersects the mantle Hugoniot to achieve both pressure and particle velocity continuity across the boundary. The impedance-match velocity of the mantle is greater than the particle velocity of the shock within the core before release. Therefore, the acceleration of the mantle leads to a release wave that propagates back into the core region. Due to the nearly perfect alignment of the target and impactor along their movement vector in head-on impacts, the ring structure might be caused by the release wave propagating towards the centre of the core.

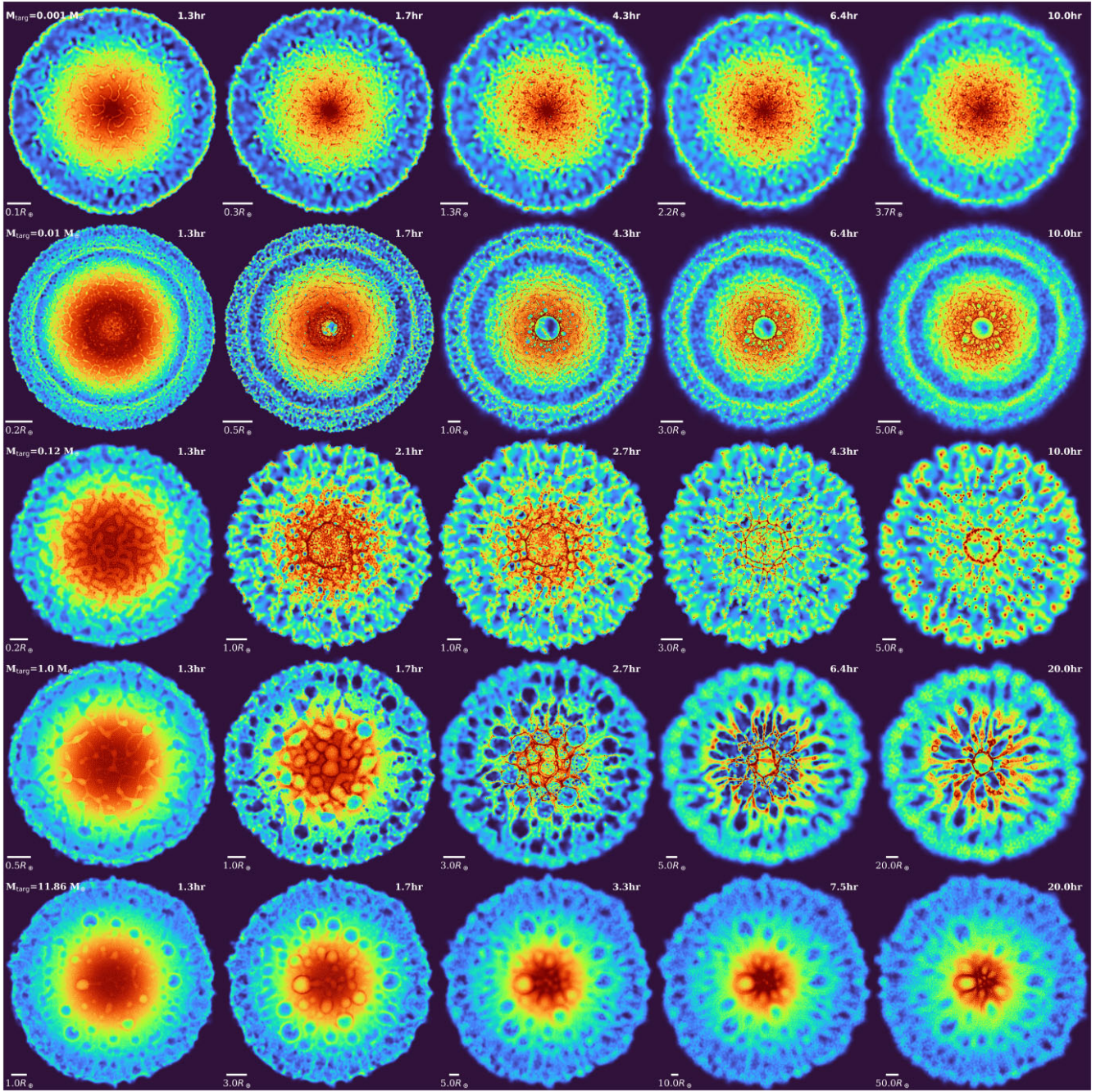

**Figure 3.** Time sequences of SPH impact simulations, showing the density of core material in the *y*-*z* plane for impacts with different target masses. The impacts were set with the target and impactor moving towards each other along the *x*-axis and so the sub-panels provide a view along the impact direction. All impacts shown are equal-mass head-on impacts; masses are displayed in the top left corner of each panel. The impact velocities correspond to those shown in the bottom row of Fig. 2 where the velocity is just high enough that the fragmentary disintegration start to happen. Colours represent densities, with denser regions appearing red and less dense regions appearing blue.

For a target mass of $0.001\,M_\oplus$, the impact velocity and initial contact shock are much weaker, and may not be able to create a strong enough release wave back into the core region, and therefore no ring structure appears. As the target mass increases towards 12 $M_{\oplus}$, higher impact energies are required for super-catastrophic impacts, which may make it harder to preserve the ring structure due to the extreme shock experienced.

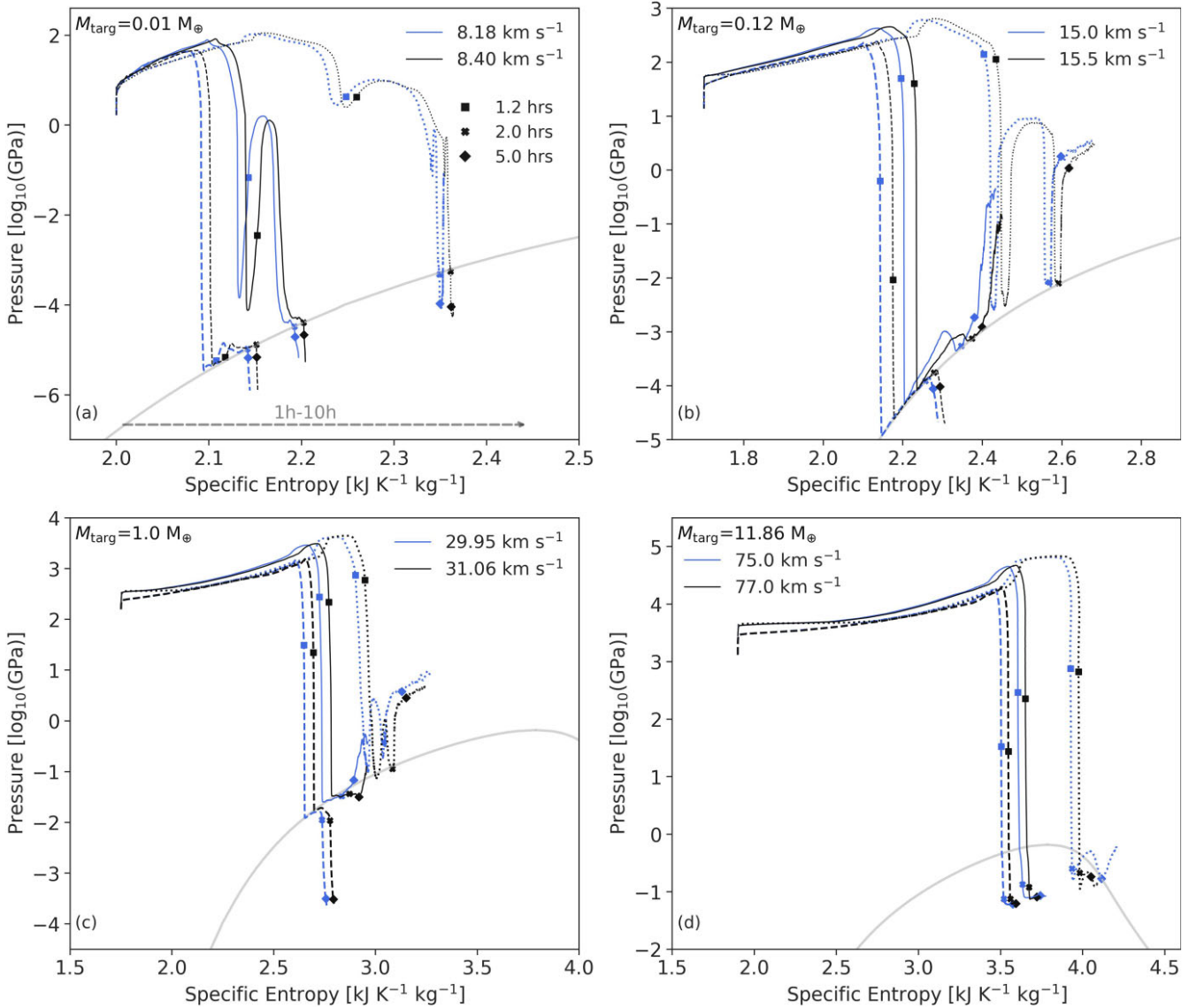

**Figure 4.** The thermodynamic evolution of core material between 1 and 10 h shown in entropy–pressure phase space. Solid lines represent the median of the entropy and pressure of tracked core particles. Dashed lines and dotted lines represent the 10th and 90th percentiles of the core entropy and pressure distribution. Therefore, each point on the lines represents certain percentiles of the tracked entropy distribution and pressure distribution at a specific time step. The bold grey lines show a fraction of the vapour dome for the iron. Impacts with velocities just below and above the critical velocity for fragmentary disintegration are shown by blue and black, respectively, for each target mass. The lines should be read from left to right to understand the thermodynamic evolution of core material with time from 1 to 10 h. The square, cross, and diamond symbols mark the state at 1.2, 2.0, and 5 h, respectively.

Giant impacts are chaotic and energetic processes, and it would be rare to see a ring structure in reality. It is likely that the extreme shear in the impact would lead to the growth of instabilities that nucleate at length-scales below the resolution of our simulations and may be actively suppressed by SPH (Hopkins 2015). These instabilities would break up the rings and filaments we see in our simulations and create less ordered structures. For this reason, it is important not to overinterpret the spatial mass distribution found in simulations of nearly perfectly aligned, head-on impacts, as the artificial structure formed likely does not approximate the real mass distribution produced by a giant impact. However, it is noteworthy that the spatial distribution of core material for impacts with different target masses becomes greatly varied as we approach the super-catastrophic regime.

### 3.3 Comparison of thermal pathways for collisions with varying target masses

Fig. 4 shows the thermodynamic changes undergone by core material in the entropy–pressure phase space during collisions with different target masses. The points on the curves with different line styles represent the 10th, 50th, and 90th percentiles of the pressure and entropy distributions of the tracked core particles, calculated sepa-rately at each time step. These 'paths'[4] provide a general overview of the core material's thermodynamic history. To complement this general overview, Fig. 5 highlights the actual thermodynamic path of three selected core particles from each target mass group, revealing more detailed impact history of the core material. In addition, to demonstrate the strength of re-shocks after the initial shock, in Fig. 6, we show how $\rho/\rho_{\max}$ ($\rho$ is the core material density and $\rho_{max}$ is the maximum core density reached during the initial shock) varies in density–entropy phase space.

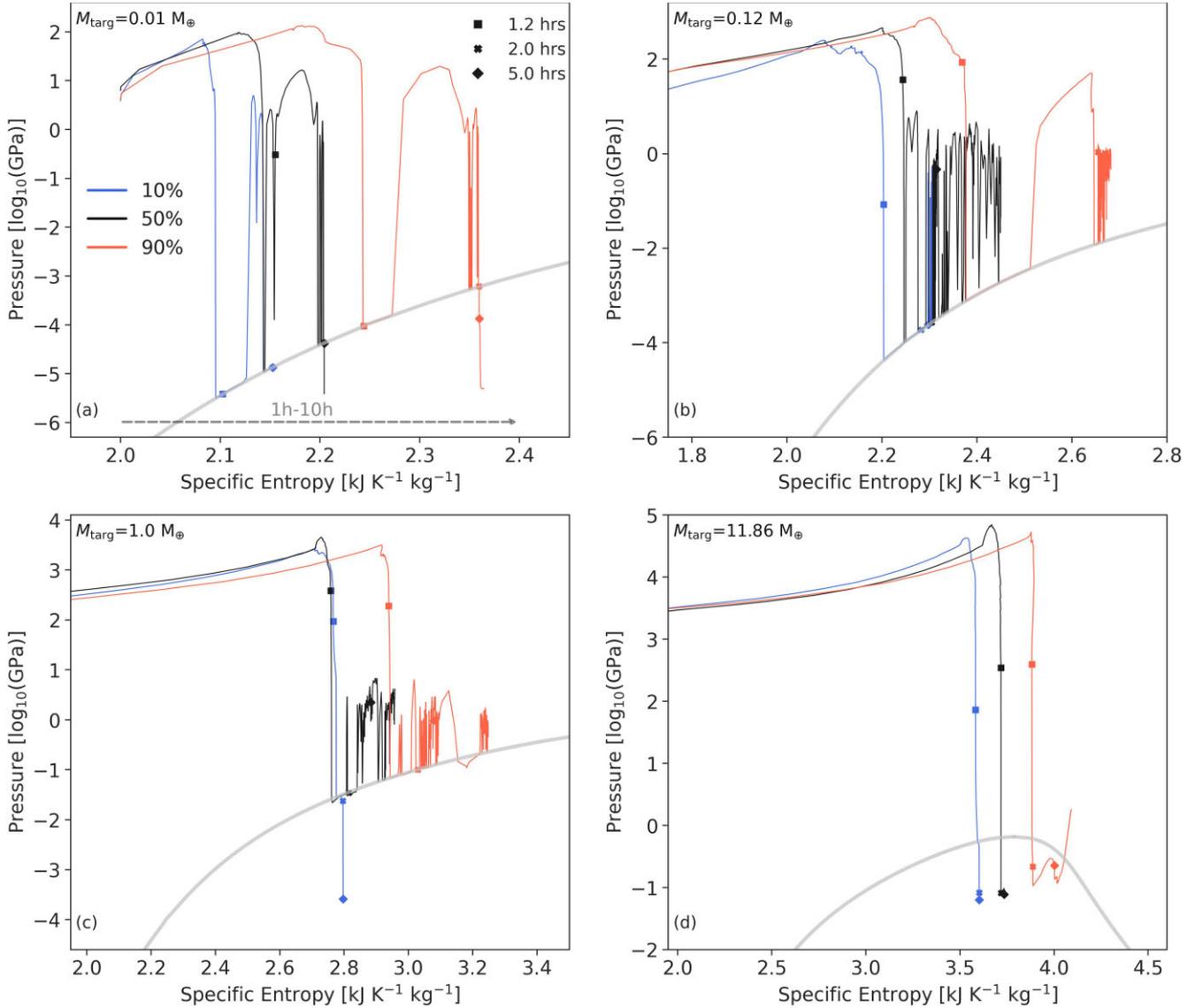

**Figure 5.** The thermodynamic evolution of three individual core particles during the first 1–10 h of the impact in the entropy and pressure phase diagram. For each target mass group, we display three particles whose entropies correspond to the 10th, 50th, and 90th percentiles of the core particle entropy distribution at the 10-h simulation snapshot. The bold grey line represents a portion of the vapour dome. The impact velocities are 8.40, 15.5, 31.06, and 77 km s$^{-1}$, which are just above the critical impact velocities to trigger fragmentary disintegration and corresponding to the bottom row of Fig. 2. The square, cross, and diamond symbols mark the state at 1.2, 2.0 and 5 h, respectively.

Fig. 4 compares impacts with velocities slightly above and below the critical velocity required to trigger fragmentary disintegration and their corresponding thermodynamic evolution paths. Within each target mass group, the thermodynamic histories of impacts with and without fragmentary disintegration are generally similar. This suggests that fragmentary disintegration is likely caused by the accumulated effects of thermodynamic changes and gravitational evolution rather than a distinct difference in thermodynamic evolution.

The thermodynamic paths in Figs 4 and 5 illustrate that during the initial collision the core material is significantly compressed and shocked, reaching very high pressures. This highly compressed state is unloaded by a release wave that propagates through the colliding bodies, adiabatically expanding to pressures that approach or hit the vapour dome. After this initial shock and release, the core material undergoes multiple re-shocks, likely caused by the fallback of ejected material and gravitational re-equilibration, leading to continued increases in entropy. In the thermodynamic paths in Figs 4 and 5, the core material interacts with the vapour dome in two different ways: 'penetrating' through the vapour dome and continuing to expand, resulting in relatively low final pressures; or having pressure–entropy paths oscillating around the liquid side of the vapour dome or following along the vapour dome before often getting re-shocked, resulting in moderate final pressures.

---

[4]Strictly, not real paths as they represent the general trend of core material but not specific particles.

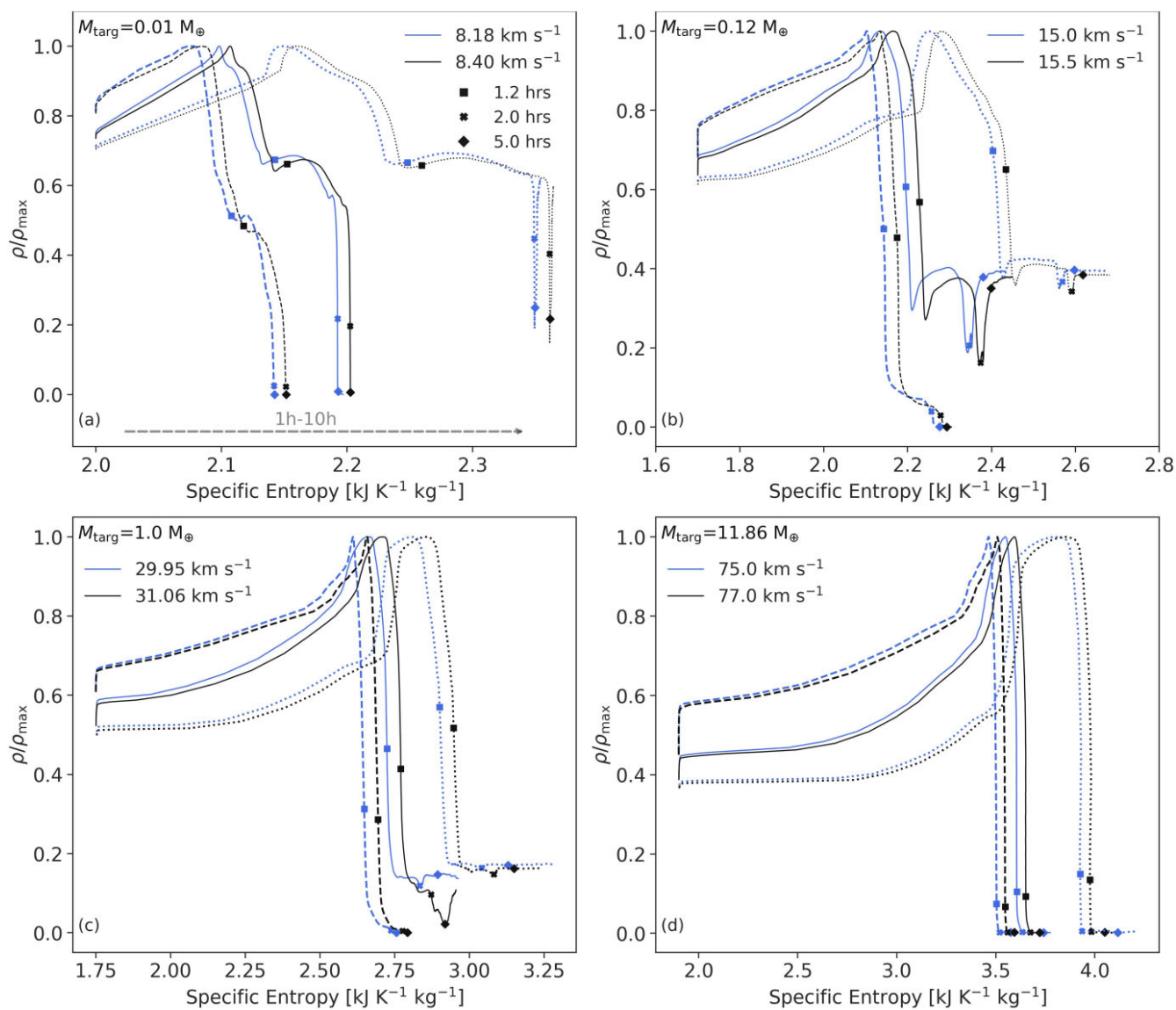

**Figure 6.** The thermodynamic evolution of core material between 1 and 10 h shown in entropy–density phase space. Densities are represent by the ratio between density at a given time step and the maximum density reached during the initial maximum compression state. Solid lines represent the median of the entropy and density of tracked core particles along time. Dashed lines and dotted lines represent the 10th and 90th percentiles of the separate tracked core entropy and density distributions. Impacts with velocities just below and above the critical velocity for fragmentary disintegration are shown by blue and black, respectively for each target mass. The lines should be read from left to right to understand the thermodynamic evolution of core material with time from 1 to 10 h. The square, cross, and diamond symbols mark the state at 1.2, 2.0, and 5 h, respectively.

Which of these behaviours a particle follows depends on the relative motion of surrounding particles. When a parcel of material hits the vapour dome, further reduction in pressure leads to a significant change in density, as $dP/d\rho$ (pressure gradient with respect to density) sharply increases within the vapour dome (Stewart et al. 2020). If a particle is unable to expand – i.e. unable to reduce its density – due to being surrounded by other particles, it will be trapped and follow along the vapour dome (oscillate repeatedly on the liquid side of the vapour curve as shown for 50th percentile paths of impacts at target masses of 0.12 and 1.0 $M_\oplus$) in pressure–entropy space. However, if a particle can expand – such as those 'penetrating' into the vapour dome – it will cross the boundary and follow the Riemann integration to very low pressures and high velocities.

The 10th percentile thermodynamic 'paths' in Fig. 4, are more likely to penetrate into the vapour dome. Often, it represents outer layer particles far away from the impact site and antipode near the $y$–$z$ equatorial plane of the core region (see blue points in second row of Fig. 17). These particles experience a weaker initial shock since they are farther away from the impact site and thus have lower entropy gain due to the initial shock. During the highest compression phase, these particles are expelled to large distance and later ejected from the system. As a result, they have a lower probability of experiencing fallback re-shocks and end up with minimal further entropy gain. On the other hand, the 90th percentile line represents particles near and along the impact velocity vector, and these particles experience the strongest shock during a head-on impact and, consequently, have a very large initial entropy gain (pink points in second row of Fig. 17).

Core material in impacts with different target masses show different interactions with the vapour dome. As target mass increases from 0.01 to 0.12 $M_\oplus$, core particles' thermodynamic paths tend to oscillate more frequently around the liquid side of the vapour dome, possibly due to the multiple reverberating shocks. At 0.01 $M_\oplus$ (Fig. 4a), the core experiences a moderate secondary shock (panel a in Fig. 6) after being released from the initial compression state. Later, the final state reaches lower pressure than impacts at higher target masses due to the weak gravitational field. As a consequence, most particles tend to penetrate into the vapour dome and expand freely, ending in a mixed liquid and vapour state. In contrast, impacts at 0.12 $M_\oplus$ (Fig. 4b) have a relatively strong secondary shock (panel b in Fig. 6), and do not decompress to as low a pressure. Therefore, more particles were trapped along the liquid side of the vapour dome and went through more complex re-shock and release processes. These re-shocks, likely caused by the re-accretion of the ejected material, prevents particles from entering the vapour dome and eventually results in more particles remaining fully liquid.

The gravitational potential wells of systems with target masses of 0.001 and 0.01 $M_\oplus$ are less deep, making it difficult to re-accrete the ejected material and, therefore, lead to fewer re-shocks. As the target mass increases, the gravitational potential well of the system becomes deeper, making it harder to expel material from the centre. This explains why more re-accretion occurs, leading to additional re-shocks. These subsequent shocks significantly affect the system's final thermal and phase state. Notably, an impact on a target mass of 0.01 $M_\oplus$ at 8.18 km s$^{-1}$ results in a higher core vapourization fraction than an impact on a target mass of 0.12 $M_\oplus$ at 15.0 km s$^{-1}$. More frequent re-shocks happened at 0.12 $M_\oplus$, resulting in more core particles left on the liquid side above the vapour dome and thus lower core vapourization fraction.[5]

Impacts at 1.0 $M_\oplus$ have higher initial entropy gain in the core material, shown in Fig. 4(c), since the initial shock intensifies with rising impact velocity. However, compared to the 0.12 $M_\oplus$ impacts, the subsequent re-shocks for 1.0 $M_\oplus$ are less energetic compared to the initial shock. As target planets become more massive, the pressure and density released from the initial compression state is also higher. At 1.0 $M_\oplus$, the first released pressure ranges from $10^{-2}$ to $10^{-1}$ GPa, while for 0.12 $M_\oplus$ impacts, it is around $10^{-5}$ to $10^{-3}$ GPa. The higher pressure and density of the material make it harder to compress during subsequent shocks, leading to less significant re-shocks.

It's noteworthy that during an impact at 31.1 km s$^{-1}$, slightly above the critical velocity necessary for triggering fragmentary disintegration at 1.0 $M_\oplus$ (as illustrated by the black solid line in Fig. 4c), some core particles initially penetrate into the vapour dome before being shocked back to the liquid side. This suggests that even at a 1.0 $M_\oplus$ target mass, the initial shock is potent enough for some core material to overcome the system's gravitational potential and expand quickly into the vapour dome, making re-accretion more difficult. As the impact velocity increases, the initial compression becomes so powerful that it can expel more material from the system resulting in fewer re-shocks.

At a target mass of around 12 $M_\oplus$ (Fig. 4d), the system has a substantial gravitational well. However, the impact velocity close to the super-catastrophic regime is high enough that the initial collision shock is strong enough to expel all the material away from the impact site. In this situation, fewer re-shocks occur, and most particles can still expand when they encounter the vapour dome. Only a few particles experience re-shocks and end up on the pure-vapour side of the dome (Fig. 5d rightmost thermal path).

In Fig. 7, we display the state of core and mantle particles in the pressure and entropy phase space at 10 h for various target masses. Immediately following a collision, impacts at different target masses display unique thermal states of the core and mantle.

As the target masses increase, the core particles follow an 'inside–outside–inside' trend with respect to the vapour dome. At low target masses (0.01 and 0.001 $M_\oplus$), most core particles are within the vapour dome, indicating the beginning of core vapourization. At intermediate target masses (0.12 and 1.0 $M_\oplus$), core particles predominantly reside above the vapour dome, on the liquid side, due to their higher pressures, showing less vapourization and more liquid formation. At a higher target mass of 5.78 $M_\oplus$, core particles again start to be trapped inside the vapour dome, experiencing more vapourization. Finally, at 11.86 $M_\oplus$, most core particles are left within the vapour dome or fully vapourized, reflecting extreme vapourization due to the high-impact energies. This evolving trend highlights the different mechanisms involved during catastrophic impacts at varying target masses. The levels of shock and gravity in head-on impacts vary significantly with the target mass.

### 3.4 Mass and iron mass fraction

The mass and iron mass fraction of the largest remnant ($M_{lr}/M_{tot}$ and $M_{Fe}/M_{lr}$, respectively) behave differently for impacts with different target masses. Fig. 8 shows the mass and iron mass fraction of the largest remnant against the specific impact energy relative to the catastrophic disruption criteria for equal-mass head-on impact simulations. Each symbol represents an impact simulation, with different colours indicating different target masses. A list of simulations involved in the figure can be found in Table A1.

As impact energies increase, the mass (left panel of Fig. 8) of the largest remnant decreases. When $M_{lr}/M_{tot}$ drops below 0.7, representing the start of significant core material loss in the system, and moving towards 0.5 to enter the catastrophic impact regime, the slope of this trend varies with target planet masses, suggesting different physical mechanisms start to dominate differently during impact processing. When the impact energy exceeds the catastrophic disruption criteria ($Q_R/Q_{RD}^*$ greater than one), the normalized mass of the largest remnant for different target masses starts to diverge towards higher energies. The decrease in the mass of the largest remnant from impacts with target masses above 0.1 $M_\oplus$ is steeper than those with target masses of 0.01 and 0.001 $M_\oplus$. This suggests that for impacts with smaller target planet masses, it becomes more difficult to eject material for the same normalized energy.

At target masses of 0.01 and 0.001 $M_\oplus$, even though the gravitational potential of the system is low in magnitude, the impact shock is also comparably weaker, with the highest impact velocities before the onset of fragmentary disintegration being only about 5 and 8 km s$^{-1}$. Thus, the erosion efficiency for these low target mass impacts is lower. Shock-driven mass-loss for these low target mass impacts is not as efficient as for larger target masses. For targets with masses above 0.1 $M_\oplus$, the gravitational potential of the system increases in magnitude, but the impact shock also becomes significantly stronger, the impact velocities before the onset of fragmentary disintegration can be as high as 15 to 75 km s$^{-1}$, leading to more efficient shock-induced mass ejection. As $Q_R/Q_{RD}^*$ continues to increase beyond 1.5, for target masses of 0.01 and 0.001 $M_\oplus$, the mass of the largest remnant switches from a linear to a power-law trend with impact energy (Fig. 8, left panel). Even though impact velocities are increasing and the initial shock should be stronger, it actually becomes harder to lose mass.

---

[5]Caveat: no long-term re-accumulation is considered here when calculating vapourization fraction.

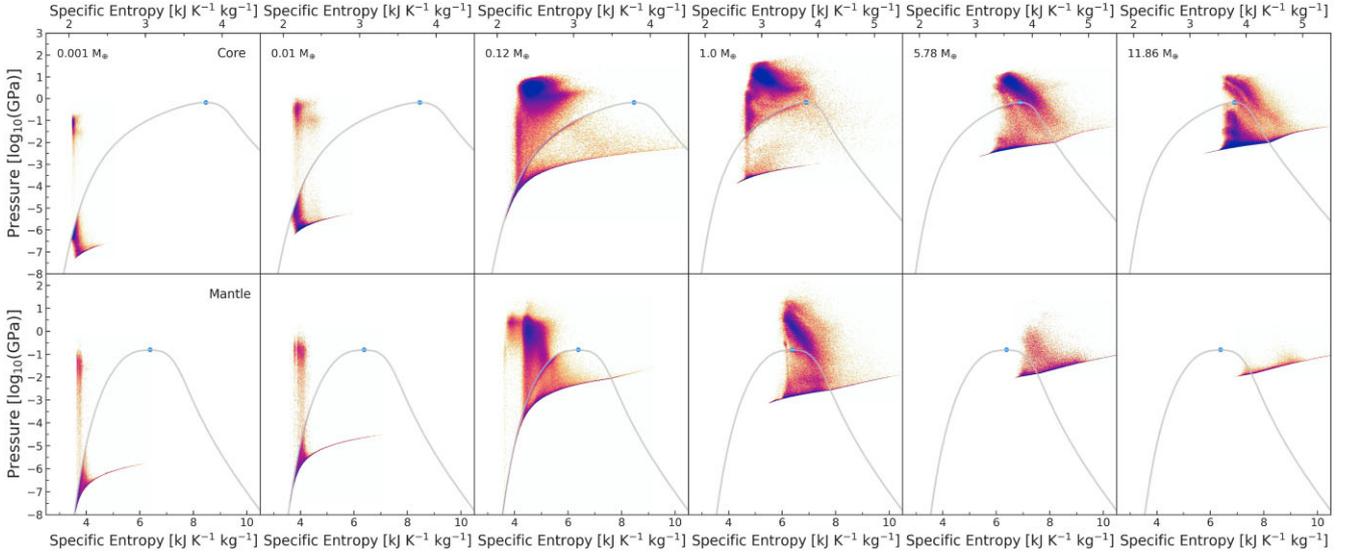

**Figure 7.** The thermodynamic states of core particles (top row) and mantle particles (bottom row) at 10 h in simulations for various target masses. The velocities of the collisions for each target mass are just below the critical velocities necessary to initiate fragmentary disintegration. The solid grey curve depicts the vapour dome, and the blue dots correspond to the critical points for the iron and forsterite EoS used in the simulations. The colour shading's intensity indicates the particle saturation in that phase space region, with bluer colours indicating a greater numbers of particles. Particles at low pressures are distributed close to a single pressure-entropy curve, stemming from particles reaching the density floor.

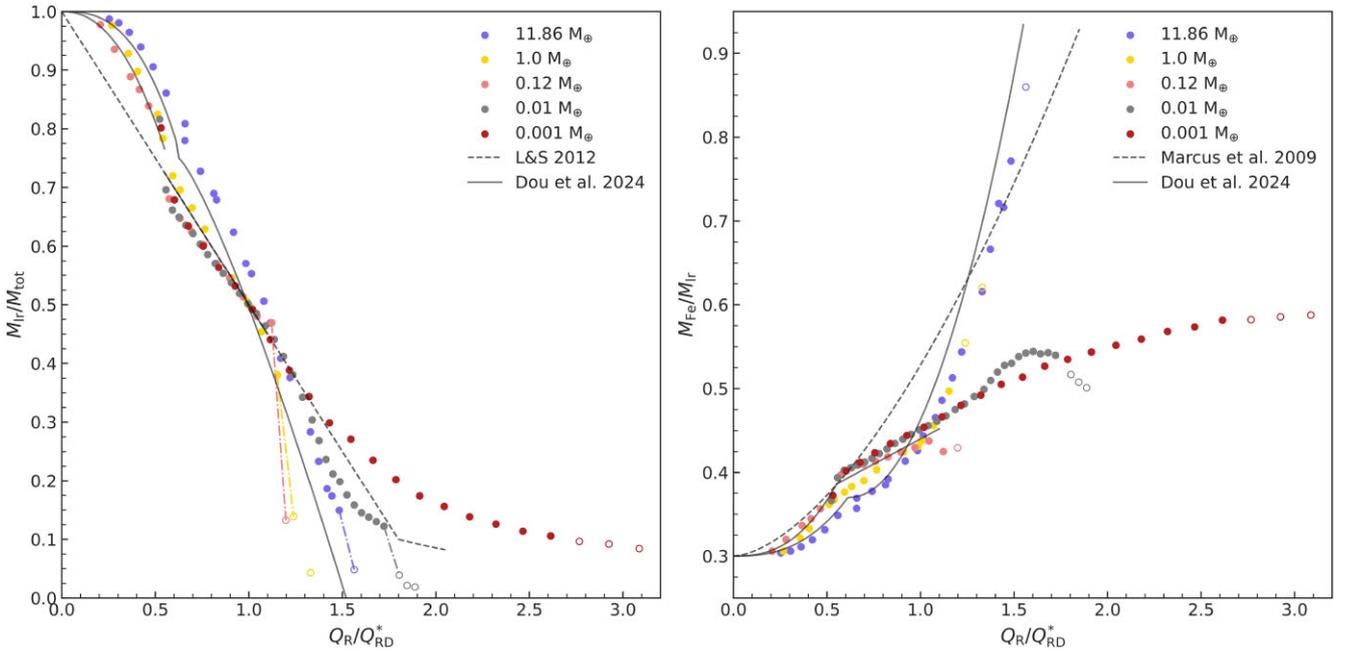

**Figure 8.** Normalized mass (left panel) and iron mass fraction (right panel) of the largest post-collision remnant, plotted against the normalized impact energy $Q_R/Q_{RD}^*$ for equal mass head-on collisions at different target masses. The dashed lines in the left panel represent the universal law for mass of the largest remnant from Leinhardt & Stewart (2012), where there is a break in the slope at $Q_R/Q_{RD}^*$ equal to 1.8, above which is the super-catastrophic disruption power law. The dashed curve in the right panel shows the scaled down version of the empirically fitted scaling law for iron fraction of the largest remnant from Marcus et al. (2009) and updated by Carter et al. (2018). The solid lines represent the scaling laws for mass and iron fraction of the largest remnant from Dou et al. (2024). Unfilled symbols indicate simulations where the largest remnant has undergone fragmentary disintegration. The simulation symbols right before (filled symbols) and after (unfilled symbols) disintegration as shown in Fig. 2 are linked with dash–dotted lines in the left panel.

Interestingly, as shown in Fig. 8 (right panel), the iron mass fraction of impacts with target mass 0.01 $M_\oplus$ shows a different increasing trend above $Q_R/Q_{RD}^*$ around 1.4. As discussed in Section 3.3, due to the low gravity and fewer re-shocks experienced by impacts in this target mass regime, much of the core material remains at lower pressure in the vapour dome and is partially vapourized. Impacts with a target mass of $0.01\,M_\oplus$ have a higher iron mass fraction than those with $0.001\,M_\oplus$ when $Q_R/Q_{RD}^*$ is between 1.4 and 1.8 possibly due to higher core vapourization fraction. This result aligns with the conclusion by Dou et al. (2024) that core vapourization enhances mantle stripping efficiency and thus leads to a higher final iron mass fraction of the largest remnant. Above $Q_R/Q_{RD}^*$ of 1.8, the largest remnants of $0.01\,M_\oplus$ impacts start to fragment, leading to a decrease in the iron mass fraction.

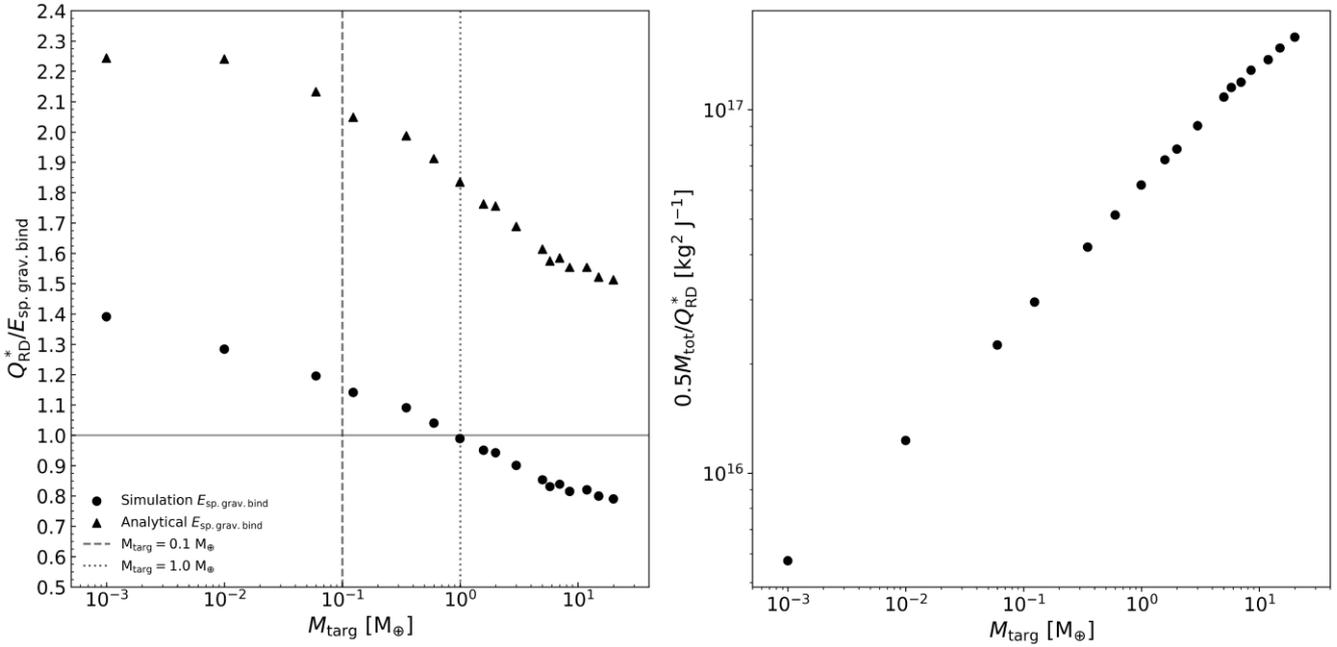

**Figure 9.** Left panel: the ratio between the catastrophic disruption impact energy ($Q_{RD}^*$) and the specific gravitational binding energy of the whole system (circles). Triangle symbols represent the ratio with gravitational binding energy, calculated using an analytical solution where each planet's potential energy is calculated by $3/5\ GM_{planet}^2/R_{planet}$. Right panel: The mass that is expelled per unit of specific impact energy at the empirical catastrophic disruption threshold is presented for different target mass groups. Data for target masses not presented in this work were derived from Dou et al. (2024).

Fig. 9 illustrates how shock and gravity behave differently at different target masses and how the efficiency of energy deposition into the ejected mass varies across different target mass regimes. The left panel shows the ratio between specific impact energy needed to remove half of the system's mass and each system's gravitational binding energy. The ratio decreases with increasing target mass, indicating that more impact energy compared to gravitational binding energy is needed to disturb lower target mass systems. This implies that the magnitude of shock is relatively lower compared to the system's gravity. The right panel demonstrates that the efficiency with which energy is deposited into the ejected mass increases for higher target mass systems. For impacts with higher target masses, more mass can get ejected per unit of impact energy compared to lower target mass impacts. Lower target mass systems tend to have lower deposition efficiency, consistent with the mass ratio slope shown in Fig. 8.

Shock thermodynamics, gravity, and core vapourization influence the final properties of the largest remnant differently at varying target masses and impact energies. We have found that the difference in thermodynamic histories and energy deposition efficiencies are particularly apparent for head-on impacts. It is challenging to derive scaling laws or draw conclusions that apply consistently across a wide range of target mass regimes when many mechanisms are involved and the balance between them is changing continuously. As clearly shown in Fig. 8, two of the most crucial properties of largest remnants, the mass and core mass fraction, change significantly when impacts enter the catastrophic impact regime. Therefore, although they give useful insight into the processes at play, head-on impacts may not be the best configuration for studying the statistical results of giant impacts. Instead, more computational resources should be allocated to the more common case of oblique impacts.

### 3.5 Impacts at $1.0\,M_\oplus$

In the previous sections, we compared how impacts with different target masses behave differently at high-impact energies, leading to catastrophic and super-catastrophic impacts. In the following sections, we focus on impacts of $1.0\,M_\oplus$ targets with impact velocities ranging from the mutual escape velocity to extremely high velocities to demonstrate the evolution of thermal properties and energy across a large impact energy range. In combination with results from previous sections, we suggest the fragmentary disintegration could be due to the cumulative interaction of core material with the liquid side of the vapour dome.

#### 3.5.1 Thermal processing

Fig. 10 illustrates the thermodynamic state of all core and mantle particles in the simulations for $1.0\,M_\oplus$ head-on impacts after 25 h of simulation time, spanning low-to-high-impact velocities from left to right. As the impact velocity increases from one to roughly three times the mutual escape velocity, both core and mantle material tend to have a narrower entropy range. At low-impact velocities, most of the core melts, with only a few particles reaching the vapour side of the dome directly. When the impact velocity increases to $22.19\,\mathrm{km\,s^{-1}}$, a new group of core particles appears: some low entropy core particles reach sufficiently low pressures to begin to intersect the liquid side of the vapour dome, with a few ending up within the vapour dome. Above approximately $30\,\mathrm{km\,s^{-1}}$, fewer particles are shocked to the vapour side of the dome directly; instead, an increasing number of particles becomes trapped on the liquid side.

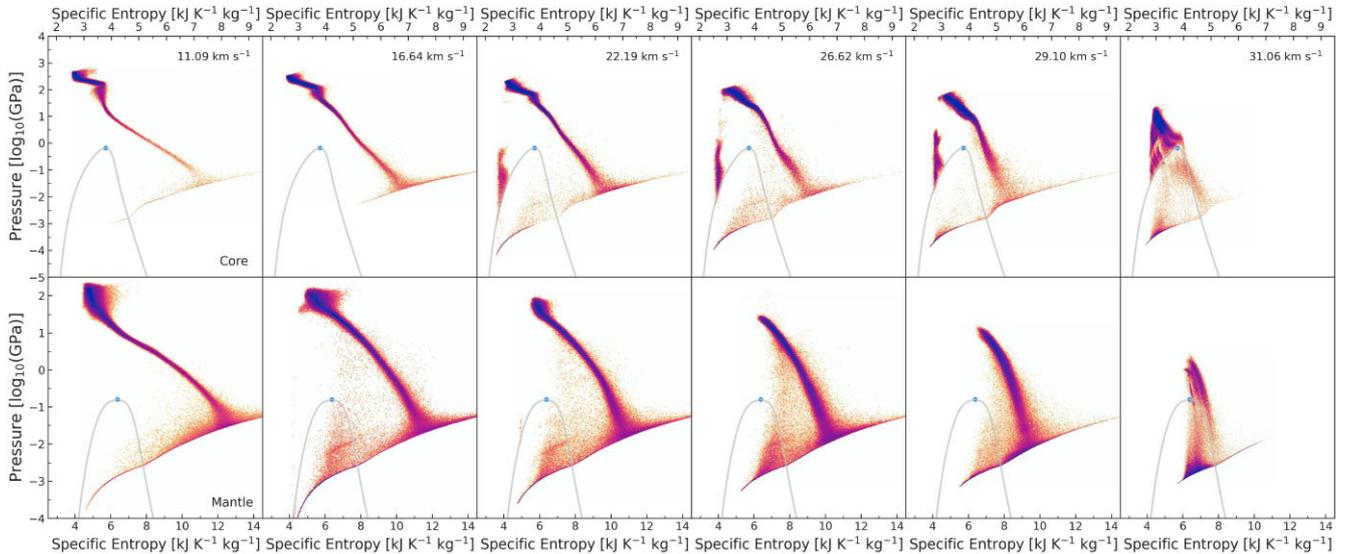

**Figure 10.** The thermodynamic states (in entropy and pressure space) of core particles (in the top row) and mantle particles (in the bottom row) 25 h after the impacts at different velocities for a target mass of $1.0\,M_\oplus$. All the impacts are equal-mass head-on impacts. The mutual escape velocity is $11.09\,\mathrm{km\,s^{-1}}$. The solid grey curve depicts the vapour dome, and the blue dots correspond to the critical points for the iron and forsterite EoS used in the simulations. Particles distributed along a curved line at low pressure have reached the density floor. The intensity of the colour shading signifies the particle density in each portion of the phase space.

Contrary to intuition, as the impact velocity increases beyond a threshold value ($\sim 28\,\mathrm{km\,s^{-1}}$ for $1.0\,M_\oplus$ target mass impacts), the system's entropy gain decreases. Fig. 11 shows the entropy and temperature gain with respect to initial states for $1.0\,M_\oplus$ target impacts as a function of impact velocity. When the impact velocity nears $28\,\mathrm{km\,s^{-1}}$, there is a sharp decline in temperature and entropy gain, aligning with the thermodynamic state depicted in Fig. 10. The temperature gain of both core and mantle particles initially increases with the rise in impact velocity. However, after peaking, it starts to decrease. In contrast, entropy gain does not exhibit a clear trend of increase or decrease until the impact velocity is sufficiently high ($\sim 28\,\mathrm{km\,s^{-1}}$), at which point it drops significantly. The range of entropy gain (the difference between 90th and 10th percentiles line) is also substantially smaller at the highest impact velocities. This pattern demonstrates how energy exchange varies with increasing impact velocities, as discussed in Section 3.5.2.

Fig. 12 illustrates the thermodynamic paths of core particles at varying impact velocities of the $1.0\,M_\oplus$ impacts. From left to right, the panels depict the evolution of the 10th, 50th, and 90th percentiles of the separate pressure and entropy distributions over time. The increases in entropy are primarily due to (1) the initial shock, which transitions kinetic energy into internal energy within the compressed body, and (2) subsequent multiple re-shocks resulting from re-accretion and gravitational re-equilibration.

Low-velocity impacts have a weak initial shock, contributing minimally to the overall entropy gain, while subsequent re-shocks account for most of the core entropy gain. Conversely, at high velocities, the initial impact contributes the bulk of the entropy gain, while secondary shocks are less frequent (due to the greater disruption and less re-accretion) and contribute less to the entropy gain. These two factors influence the system's thermodynamic state mutually, explaining why the entropy does not show a consistent increase or decrease for intermediate impact velocities, as shown in Fig. 11.

Low-velocity impacts can greatly deform the two colliding planets. This deformation leads to gravitational re-equilibration, causing substantial re-accretion and reheating but without ejecting too much material. As a result, the release pressure and density reached during decompression remains high. Consequently, most particles remain above the vapour curve on the liquid side, while a smaller number can be shocked to the vapour side of the dome directly, as shown in Fig. 10.

As impact velocities increase, the two colliding bodies can be significantly compressed. The resulting decompression reaches pressures and density low enough that material intersects the vapour dome. Due to the shape of the vapour curve, released particles that reached higher entropy (but below the critical point) in the initial shock interact with the vapour curve more readily. As the impact velocity increases, the core particles reach lower release pressures and higher entropies due to the initial shock. Consequently, more core material start to interact more frequently with the vapour curve.

As a parcel of core material hits the vapour curve, the pressure–density gradient ($\mathrm{d}P/\mathrm{d}\rho$) and the acceleration of particles for a given decrease in pressure both significantly increase, as noted by Stewart et al. (2020). Frequent interaction of SPH particles with the vapour curve may result in accumulated thermodynamic effects on the post-collision structure. In conjunction with the results presented in Section 3.3 for impacts involving various target masses, fragmentary disintegration occurred for target masses between 0.12 and $1.0\,M_\oplus$ at relatively low-impact energy, where their thermodynamic paths oscillated repeatedly or followed the vapour curve more frequently. This suggests that fragmentary disintegration could be associated with the cumulative effect of vapour curve interaction.

In the target mass range of $0.1$–$1.0\,M_\oplus$, the combination of the gravitational potential well and shock strength for head-on impacts make it more likely for core material to interact with the vapour dome. For target masses below $0.1\,M_\oplus$, such as at $0.01\,M_\oplus$, the shock is too weak to drop the pressure and density of core material low enough to interact frequently with the vapour dome. However, as the target mass increases above 1.0 up to $11.86\,M_\oplus$, the shock becomes strong enough to propel core particles directly over the vapour dome or allow them to enter and expand in the vapour dome freely, as illustrated in Fig. 7.

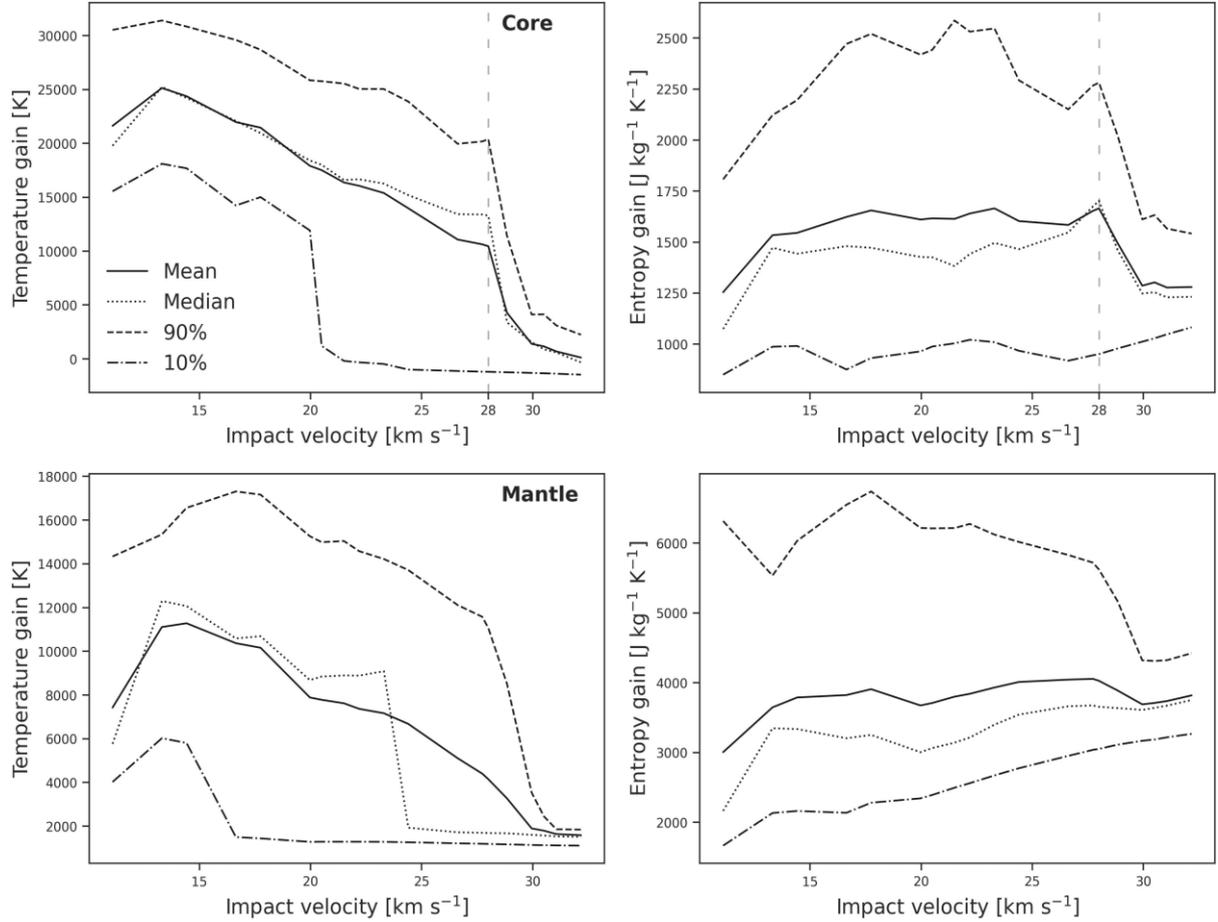

**Figure 11.** The temperature (left) and entropy (right) gains of core and mantle particles at various impact velocities for $1.0\,M_\oplus$ target mass head-on impacts. The gains in entropy and temperature are calculated based on the difference between the 0-hour and 15-hour simulation snapshots. The four different line styles represent the 10th (dash–dotted), 50th (dotted), and 90th (dashed) percentiles, and the mean (solid) of the distributions of the gains in entropy and temperature.

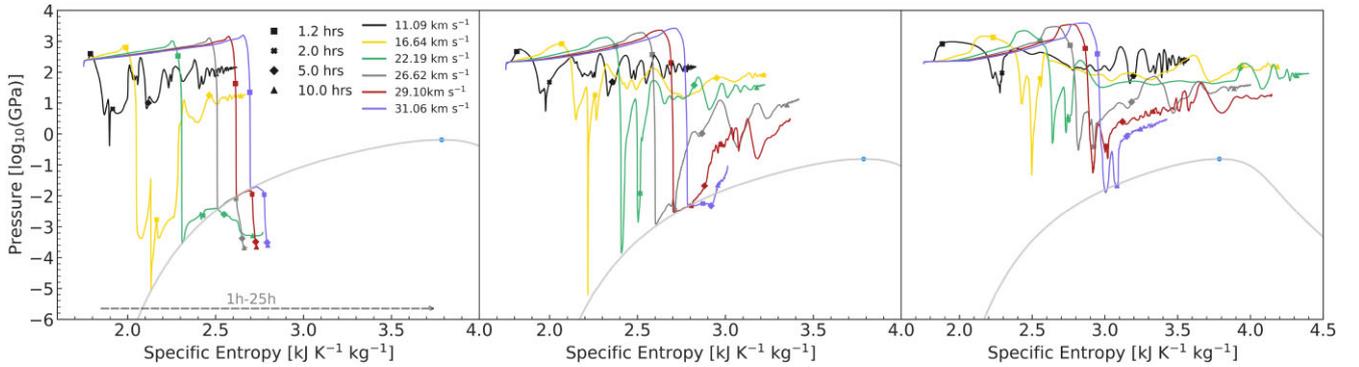

**Figure 12.** The thermodynamic evolution of core material at various impact velocities over 1–25 h in the entropy–pressure phase diagram. From left to right, the three columns represent the 10th, 50th, and 90th percentiles of the tracked properties distribution. The bold grey line represents a part of the vapour dome of the iron EoS used in the simulations. All the simulations shown here are equal-mass $1.0\,M_\oplus$ head-on impacts. The square, cross, diamond, and triangle symbols mark the state at 1.2, 2.0, 5.0, and 10.0 h, respectively.

In combination with the results from Section 3.3 for different target masses and the results here for the same target mass but different impact velocities, we show how gravity and thermodynamic interaction can significantly influence the outcomes of head-on impacts. These findings suggest two key considerations: (1) impacts for different target masses do not behave similarly, so caution should be exercised when applying a universal law to a broad range of target masses; and (2) an accurate phase boundary to represent the phase transition of material is crucial for accurately modelling the outcome of giant impacts.

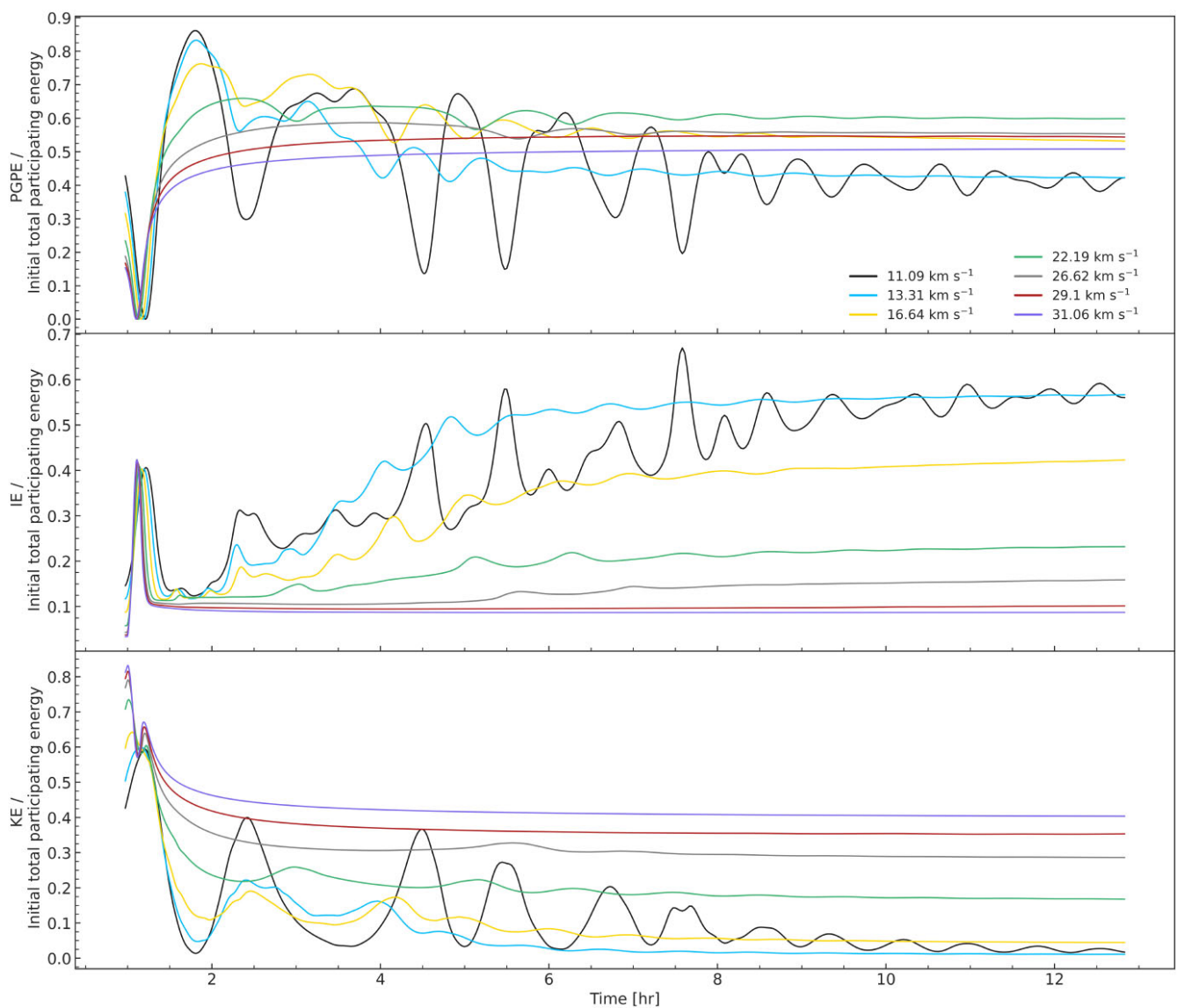

**Figure 13.** Temporal evolution of participating potential energy (top panel), internal energy (middle panel), and kinetic energy (bottom panel) of equal-mass head-on impacts for $1.0\,M_\oplus$ at various impact velocities. We show the ratio between these three energy components and the initial total participating energy in the system. Note the energies shown are for all material in the systems, not specifically the core material, as emphasized in the previous analysis. The simulations were set up such that collision between two bodies happened at around 1.0 h.

### 3.5.2 Energy exchange

The previous sections outlined the spatial distribution and thermodynamic evolution paths for various head-on impacts. Examining the temporal evolution of different energy components can provide additional insight into the impact processes occurring at various impact energies and speeds. Fig. 13 illustrates the evolution of participating potential energies, internal energies, and kinetic energies over a range of impact velocities for the target mass of $1.0\,M_\oplus$.

Before the collision, as the two bodies move towards each other, they fall into each other's gravitational pulls, causing a decrease in potential energy. This potential energy is converted into kinetic energy as the bodies speed up. As the impact begins, the potential energy continues to decrease due to the compression and deformation of the bodies. The collision shock wave then transfers energy into the bodies, leading to an increase in internal energy and a change in kinetic energy.

Shortly after the impact starts, the bodies reach a state of maximum compression, where the mass is most dense. This state corresponds to the lowest point in potential energy. Simultaneously, the shocked material begins to decompress and expand. Most of the internal energy gained from the initial impact is then converted back into kinetic and potential energy. The higher the impact speed, the sooner the system reaches a state of maximum compression.

After decompression, the exchange of energy diverges significantly. For impacts with velocities below approximately $26\,\text{km s}^{-1}$, the displaced material falls back due to the gravitational pull, reduc-ing the potential energy but not as significantly as the initial impact. As gravitational re-equilibration continues, gravity compresses the expanded fluid, and fragments fall back onto the post-impact body, causing secondary shocks. These shocks generate further heating and increase the internal energies. The recurring secondary shocks cause the evolution lines of three energy components to oscillate, demonstrating frequent energy exchange.

However, when the impact velocity increases above 26 km s$^{-1}$, the frequency and magnitude of energy exchanges decrease. The system's energy state is almost solely determined by the initial shock and barely changes afterward. This suggests that there are fewer secondary shocks and that the secondary shocks have a diminished effect on the energy state of the system. This shift in energy dynamics again underlines the complex interplay of gravity and shocks during catastrophic impacts.

In Fig. 14, we display the energy budget of impacts at varying velocities, from low to high. Each impact event results in a significant exchange between internal, kinetic, and gravitational potential energies. There are substantial energy alterations after a collision. For impacts at low velocities, a minimal amount of impact energy remains as kinetic energy. Instead, most of the impact energy transitions into internal energy (due to subsequent secondary shocks) and gravitational energy (due to the dispersal of material).

At the final state of impacts at around 25 h, as impact velocities increase, the fraction of gravitational energy initially increases and then decreases, while kinetic energy continues to rise. At an impact velocity of 31.06 km s$^{-1}$, approximately 90 per cent of the impact energy is in the form of kinetic and gravitational energy. This high percentage indicates that the system is dominated by the initial shock dynamics, with less energy available for secondary processes like re-accretion and re-shocks. The material is expelled so efficiently that it prevents significant gravitational re-equilibration, leading to a predominance of kinetic and potential energy states.

This analysis underscores the complex interplay between different forms of energy during head-on impacts. At lower velocities, the system experiences multiple phases of energy exchange, with re-accretion and re-shocks playing crucial roles. As velocities increase, the energy dynamics simplify, with the initial impact shock becoming the dominant force determining the final energy distribution.

*3.5.3 Maximum entropy increase of a single shock*

To assess the relative contributions of the initial shock and subsequent reshocks to the heating of material, we calculate the analytic maximum entropy increase that could be achieved by a single impact shock wave. Fig. 15 compares the analytic maximum entropy attainable by a single shock with the near-maximum entropies reached in SPH simulations of a 1.0 M$_\oplus$ target at various impact velocities after a simulation time of 5 h. For impact velocities below approximately 31 km s$^{-1}$, the near-maximum entropies from the SPH simulations consistently exceed the analytically calculated maximum core entropies. As the impact velocity increases above 16 km s$^{-1}$, the difference between the maximum entropy values from the SPH simulations and the impedance match calculations decreases, suggesting a diminishing contribution from reshocks to the heating and entropy gain of the system. At an impact velocity of 31.06 km s$^{-1}$, the near-maximum entropy from the SPH simulation is lower than the analytically calculated maximum entropy at the CMB, indicating that the entropy gain is predominantly affected by the initial shock at this velocity, which is consistent with the previous conclusions from Sections 3.5.1 and 3.5.2.

*3.5.4 Shock history in different regions*

In Fig. 16, we illustrate the varied thermodynamic and shock processes experienced by core particles in different regions of the target planet at 1.0 M$_\oplus$. Each point within the circle region represents the initial *x–y* position of a parcel of target core particles. In Fig. 17, we show the initial locations of core particles with different final entropy levels in their target planets for impacts at 16.64 km s$^{-1}$ (panels a and b) and 31.06 km s$^{-1}$ (panels c and d). As the impact velocity increases, the regions where particles experience the most and least entropy gain change greatly.

After 10 h of an impact, at relatively low-impact velocities (first and second rows in Fig. 16 at 11.09 and 16.64 km s$^{-1}$), the particles displaced far from the centre (shown by the yellow region in the second column) are shown by pink particles in panels (a) and (b) in Fig. 17. These outer layer particles from the hemisphere close to the impact site experience the strongest shock initially and end up with higher entropy gain. However, particles close to the impact site actually gain less entropy. Particles near the antipode (blue particles in Fig. 17a) have the lowest entropy gain, as they experience a less intense shock and fewer re-shocks.

At high-impact velocities (third and fourth rows in Fig. 16 at 22.19 and 31.06 km s$^{-1}$), the core particles that are kicked far away instead start to have lower entropy gain and lower internal energy. At 31.06 km s$^{-1}$, the outer layer core particles between the impact site and antipode (the blue particles in panels c and d in Fig. 17, near the *y–z* equatorial plane) are those that are kicked the most far away and have the lowest entropy gain. These particles correspond to the thermodynamic path in the left panel of Fig. 12 (purple line), which can directly penetrate through the liquid side of vapour dome and expand freely. In contrast, the core particles closest to the impact site and near the antipode have the largest entropy gain and higher internal energy. These are the core particles corresponding to the right panel of Fig. 12, indicating that these particles experienced the strongest shock during the initial collision and underwent a few secondary shocks later.

From low-to-high-impact velocities at the same target mass, different regions in the core experience various levels of shock and perturbation, leading to different final entropy gains. These findings underscore the significant differences in the thermodynamic histories and final states of core particles depending on their initial positions and the impact velocity. At lower velocities, as the strength of the shock is relatively low, only particles in the impact site hemi-sphere feel a significant shock and get disturbed, gaining a substantial amount of entropy. At higher velocities, the initial shock's strength predominates, leading to a more straightforward thermodynamic path with less secondary shock involvement. As a result, particles near the impact site and antipode gain the most entropy.

## 4 DISCUSSION

### 4.1 When does fragmentary disintegration happen?

In order to trigger fragmentary disintegration, several factors must be considered: target mass, impactor to target mass ratio, impact angle, and impact velocity. These factors should be configured in such a way that the core material frequently interacts with (either follows

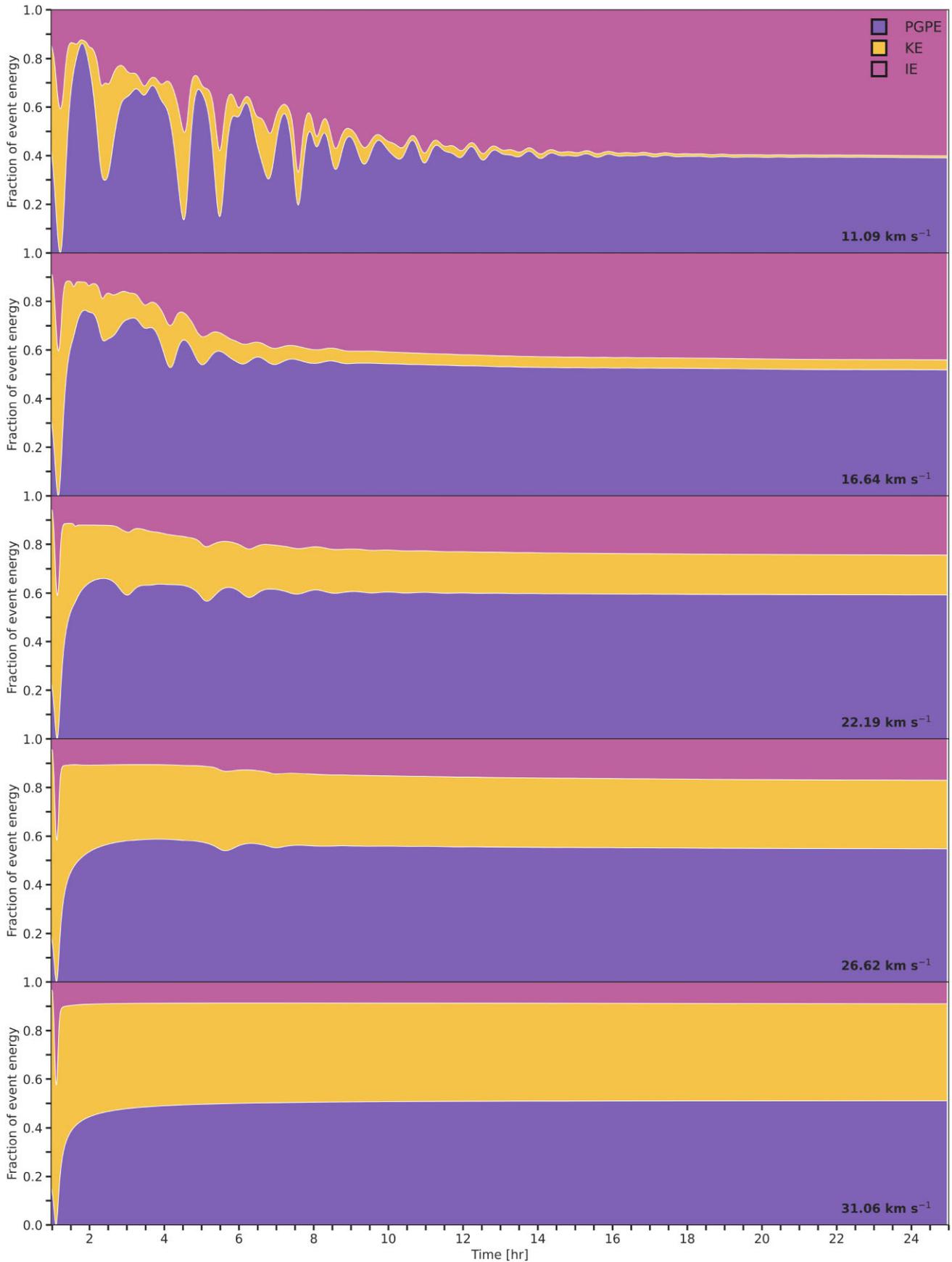

**Figure 14.** Temporal evolution of the energy budget of $1.0\,M_\oplus$ impacts from low to high-impact velocities. The energy budget is the sum of the 'participating' potential energy, $E_{pot} - E_{pot,min}$ (PGPE – purple), kinetic energy (KE – orange) and internal energy (IE – magenta). For any given time step, the height of a colour band corresponds to the fraction of that particular energy type in the total energy at the current time step.

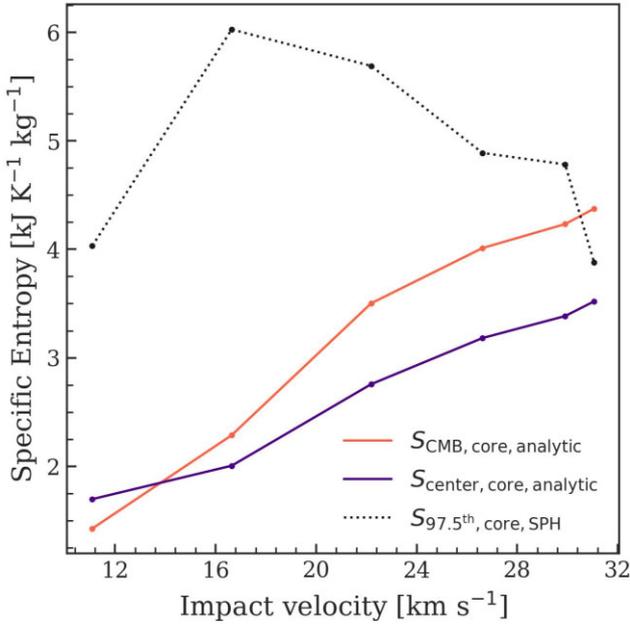

**Figure 15.** Entropy comparison between SPH simulations and semi-analytical calculations of 1.0 M$_\oplus$ head-on impacts at various impact velocities. The orange and indigo lines (points) illustrate the maximum entropies achievable for core material near the CMB and the central region, respectively, resulting from a single shock. These values were calculated using the impedance match method (see Section 2.5 for details). The dotted lines represent the near maximum (97.5th percentile of entropy distribution to exclude the effect of outliers) core entropies obtained from SPH simulations after a simulation time of 5 h. The initial entropy values for the core and mantle were set to 1750 and 3027 J K$^{-1}$ kg$^{-1}$, respectively.

along or oscillates repeatedly around) the liquid side of the vapour dome.

First, an impact should be able to disturb as much core material as possible in a planet, thereby allowing more core material to interact with the vapour curve. Therefore, the impact should be head-on or nearly so, with impactors of similar size to the target, typically with a mass ratio ($\gamma$) greater than 0.7.

Next, the target planet's mass should be at least larger than 0.01 M$_\oplus$. Below this mass, both the gravitational potential well and shock strength are low. As a result, very few re-shocks occur, leading to minimal interaction of core material with the vapour dome.

For target masses between 0.01 and 1.0 M$_\oplus$, the impactor-to-target mass ratio ($\gamma$) should be at least 0.7 to generate a sufficiently strong shock capable of triggering fragmentary disintegration. If the impactor is smaller, the resulting shock would again be too weak to induce the necessary compression and phase changes.

As the target mass increases above 1.0 M$_\oplus$, the shock becomes stronger, making it easier for the core material to vaporize. Additionally, the gravitational potential well relative to the impact energy becomes less significant (see Fig. 9). This reduced gravitational influence means that at high-impact energies, fewer re-shocks occur, resulting in core material being more likely to be directly ejected into the vapour dome without undergoing multiple compression cycles.

Moreover, as the efficiency of mass ejection increases with the target mass, fragmentary disintegration begins to occur with a smaller remaining mass in the largest remnant. For equal-mass impacts with target masses of 2.0 and 3.0 M$_\oplus$, fragmentary disintegration occurs when the largest remnant mass ($M_{lr}/M_{tot}$) is around 0.2. For target masses exceeding 7.0 M$_\oplus$, fragmentary disintegration occurs when $M_{lr}/M_{tot}$ is less than 0.1, entering the previously defined super-catastrophic disruption region.

### 4.2 What happened to the mantle?

In this study, we tracked and analysed how core material evolve at various target masses and impact energies. We concluded that different thermodynamic processing of core material results in various final states of head-on impacts.

Fig. 18 displays the mantle's thermodynamic evolution path for 1.0 M$_\oplus$ head-on impacts. Since mantle material is less dense, it is easier to be compressed and ejected. At high-impact velocities above 26 km s$^{-1}$, mantle material is either shocked into a mixed state of liquid and gas inside the vapour dome and expands freely, or it is directly vapourized to be on the gas side of the vapour dome.

Thus, during high-energy impacts, mantle material has limited interaction with the vapour dome, and their influence on thermodynamic processing could be minimal. However, it should be noted that mantle material accounts for 70 per cent of the total mass in the system, and they are all significantly disturbed and ejected during head-on impacts. As a result, mixing between core and mantle particles happens globally at high-impact energy. The SPH formulation can create artificial forces that repel one material from the other, suppressing mixing between different materials (Deng et al. 2019; Ruiz-Bonilla et al. 2022). Testing the influence of mixing is beyond the scope of this study. However, we reiterate that head-on and nearly head-on impacts, while simple in setup, involve complex mechanisms.

### 4.3 Effect of the mass ratio

In this study, all results are based on equal-mass, head-on impacts where the impactor-to-target mass ratio ($\gamma = M_{imp}/M_{targ}$) is one. Additionally, we tested a subset of head-on impacts with different mass ratios for target masses of 0.12, 1.0, and 1.58 M$_\oplus$ at $\gamma$ values of 0.1, 0.4, and 0.7. Fragmentary disintegration occurred for all three target masses at $\gamma = 0.7$, but not at $\gamma$ values of 0.1 and 0.4. The largest remnant mass ($M_{lr}/M_{tot}$) at energies just before fragmentary disintegration was approximately 40 per cent and 30 per cent for target masses of 0.12 and 1.0 M$_\oplus$ at $\gamma = 0.7$, respectively, similar to the results from equal-mass impacts. For the target mass of 1.58 M$_\oplus$, we tested $\gamma$ values of 0.5 and 0.6, and fragmentary disintegration occurred at $\gamma = 0.6$ but not at 0.5.

Next, we tested a target mass of 3.0 M$_\oplus$ at $\gamma = 0.67$ and 0.83, and a target mass of 5.0 M$_\oplus$ at $\gamma = 0.1, 0.2, 0.32, 0.4, 0.5, 0.6, 0.7$. We did not see clear evidence for fragmentary disintegration when $Q_R/Q_{RD}^*$ was less than 1.8 or $M_{lr}/M_{tot}$ was above 10 per cent. As target masses increase beyond 1.0 M$_\oplus$, material erosion efficiency also increases (Fig. 9). Large $\gamma$ values (e.g. 0.6 and 0.7) could potentially trigger fragmentary disintegration; however, fragmentary disintegration could occur when the normalized impact energy $Q_R/Q_{RD}^*$ is already above 1.8 and $M_{lr}/M_{tot}$ is below 10 per cent, which is in the previously defined super-catastrophic regime. For small $\gamma$ values, on the other hand, it is easier for the impactor itself to be vapourized and destroyed before it can significantly disturb the core of the target planet.

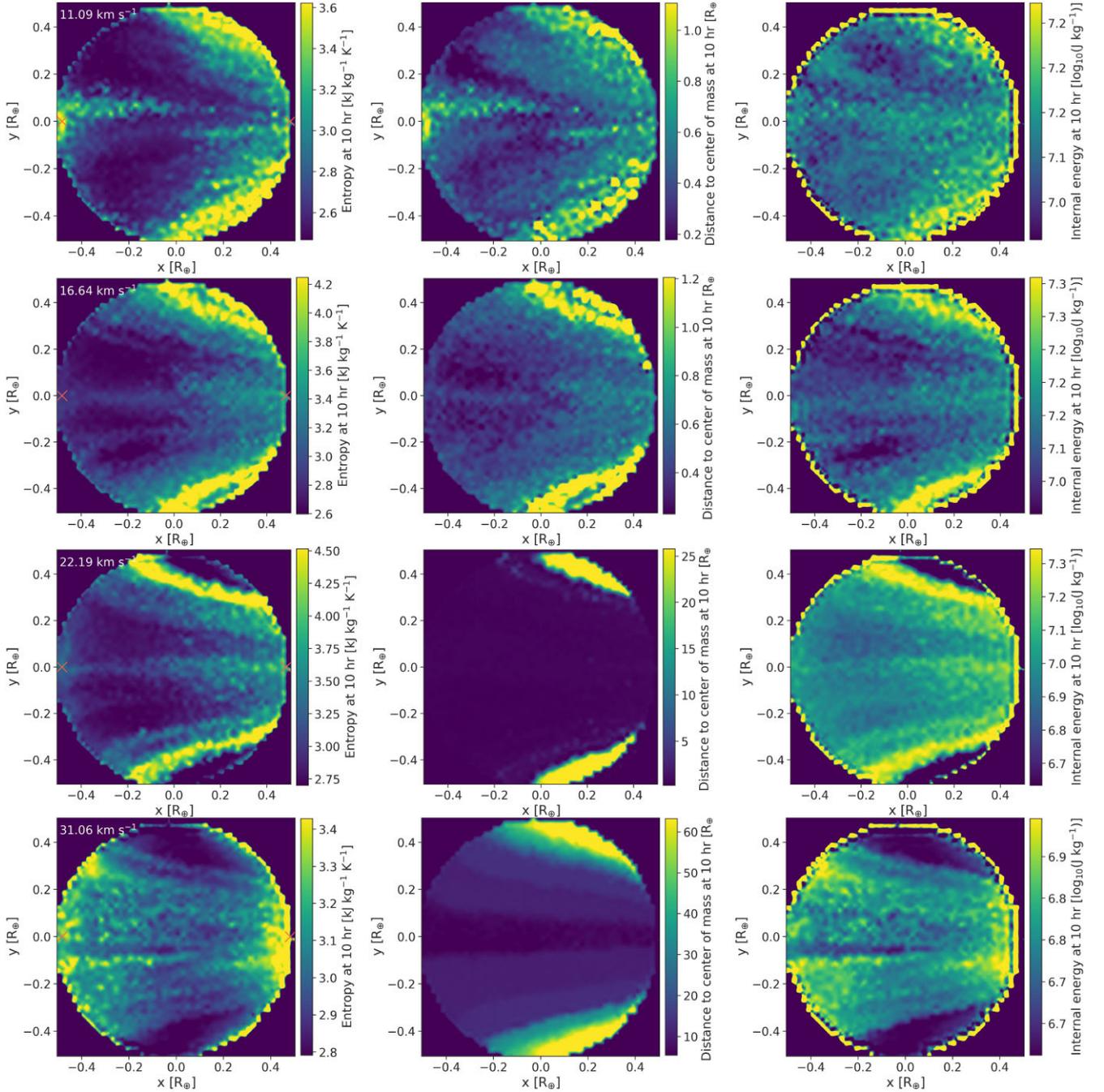

**Figure 16.** Entropy (first column), distance to the centre of mass (second column), and internal energy (third column) of target planet core particles are shown by colour at different velocities (each row) at 10 h for 1.0 $M_\oplus$ head-on impacts. We select and grid core particles initially in the target planet near the *x-y* equatorial plane with *z* axis values between −5 and 5 per cent core radius. The three properties of these selected particles at 10 h are then averaged along the *z* dimension and subsequently integrated and shown in colour in the *x* and *y* dimensions. The location of each colour cell in the plot represents the initial position of core particles, while the colour intensities show the end states of the corresponding core particles at 10 h.

### 4.4 Effect of the impact parameter

We tested a group of nearly head-on, equal-mass impacts with an impact parameter $b = 0.1$ (approximately 5.74°) for target masses of 0.12, 1.0, and 1.58 $M_\oplus$. We found that at 0.12 $M_\oplus$, fragmentary disintegration occurred when the largest remnant mass ratio ($M_{lr}/M_{tot}$) was around 40 per cent, and for target masses of 1.0 and 1.58 $M_\oplus$, fragmentary disintegration occurred when $M_{lr}/M_{tot}$ was around 30 per cent. These results were similar to those observed in equal-mass head-on impacts.

In Fig. 19, we illustrate the thermodynamic path and final state of core material from an oblique equal-mass impact with a target mass of 1.0 $M_\oplus$ and an impact angle of 30° ($b = 0.5$). Compared to the thermodynamic path of a head-on impact, an oblique impact has less interaction between the core and the liquid side of the vapour dome. In the case of an oblique impact at high velocity, a significant amount of core material is ejected directly into the vapour dome where it is in a mixed state of gas and liquid. These particles, possibly originating from the overlapping regions of the target and impactor, are often ejected from the system. The remaining core particles experience significant disturbance, entering and exiting the vapour dome without substantial repeated interaction, unlike the head-on impacts shown in Fig. 4, where particles oscillate around the vapour dome boundary along the liquid side.

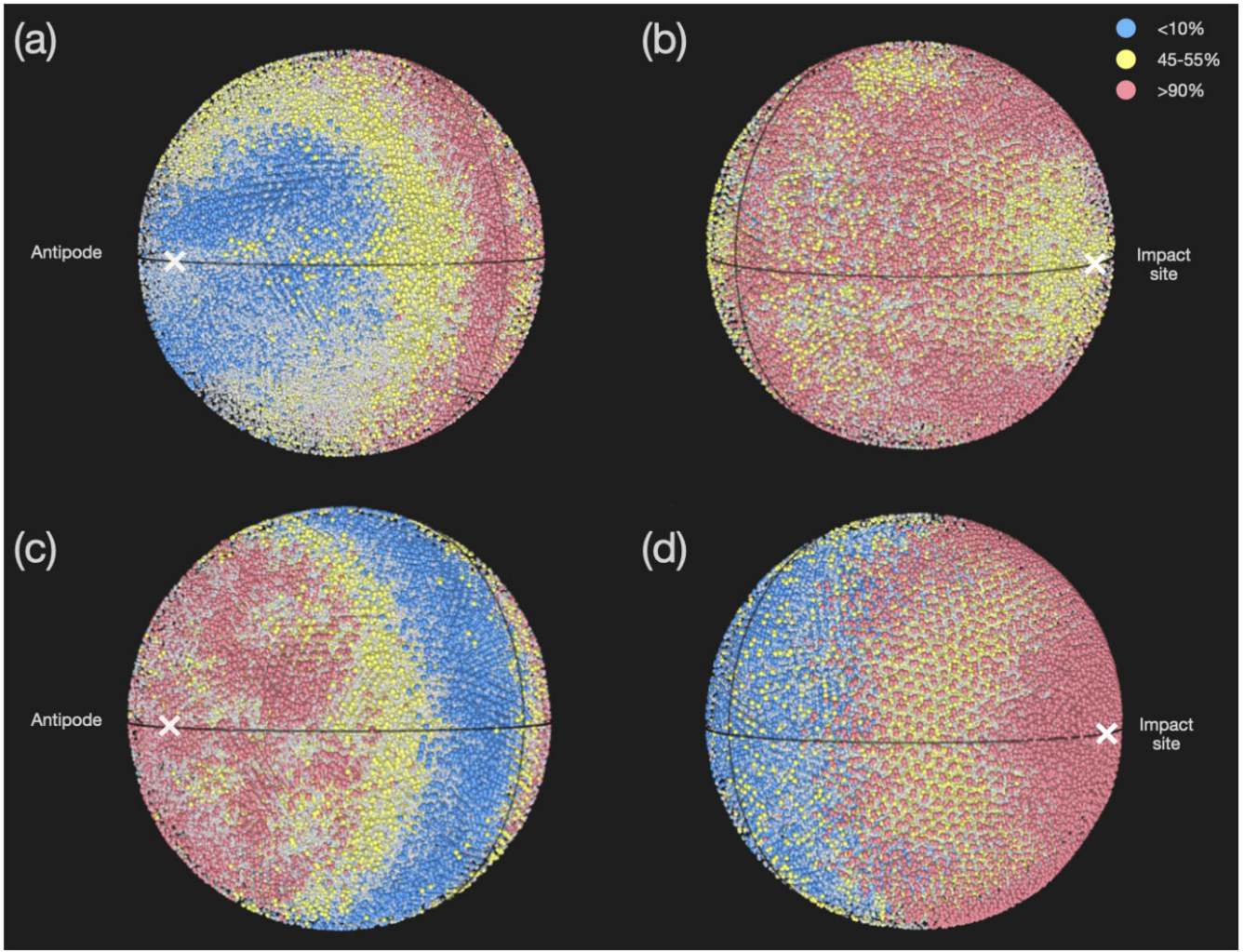

**Figure 17.** The initial core particle positions in the target planets are displayed with coloured particles showing the entropy distribution in the 10-h simulation snapshot. Blue particles indicate positions of particles whose final entropies are between the 0–10th percentiles of the final entropy distribution. Yellow and pink particles represent those between the 45–55 th and 90–100th percentiles, respectively, in the entropy distribution at 10 h. Grey particles are all the other particles that are not in the above mentioned entropy range. The top two panels (a and b) show the impact at 16.64 km s$^{-1}$, while the bottom two panels (c and d) show the impact at 31.06 km s$^{-1}$. The camera is placed around the *x*-*y* plane, with panels (a and c) having a viewing angle from the antipode and panels (b and d) having a viewing angle from the collision side. Impact site and antipode are marked with white cross. Black lines denote the *x*-*y* and *y*-*z* equatorial planes. 3D illustrations are generated using K3D-jupyter (K3D-jupyter 2015).

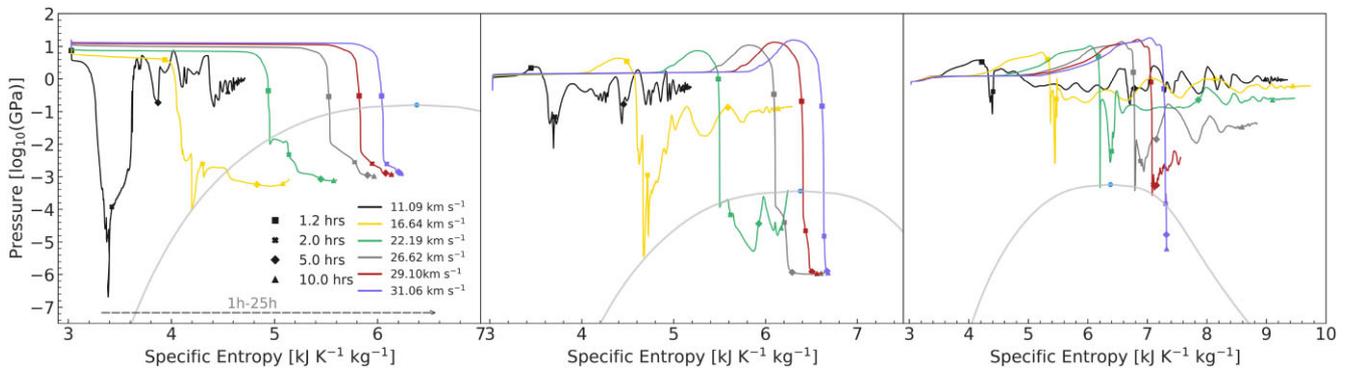

**Figure 18.** The thermodynamic evolution of mantle material at various impact velocities during 1–25 h in the entropy and pressure phase diagram. From left to right, three columns represent the 10th, 50th, and 90th percentile of the track properties distribution. The bold grey line represents partial of the vapour dome of forsterite EoS we used in the simulations. All the simulations shown here are equal-mass $1.0\,\mathrm{M_\oplus}$ head-on impacts. The square, cross, diamond, and triangle symbols mark the state at 1.2, 2.0, 5.0, and 10.0 h, respectively.

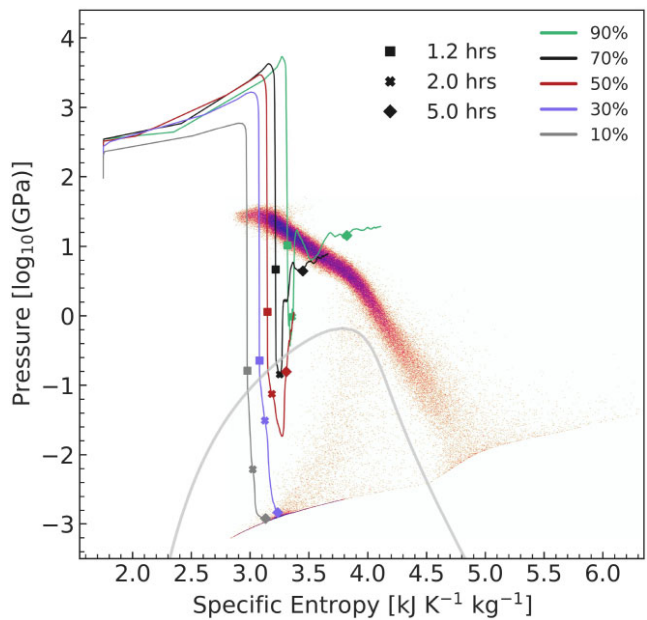

**Figure 19.** Thermodynamic evolution (1–10 h) of core material of a $1.0\,M_\oplus$ equal-mass impact at $b = 0.5$ (30°) and impact velocity $55.61\,\mathrm{km\,s^{-1}}$. Each coloured line represents a certain percentile of the separate tracked pressure and entropy distributions. The background coloured points show the thermodynamic states (in entropy and pressure space) of core particles at 10 h after the impact. The solid grey curve represents the vapour dome. Particles distributed in a curvature shape at low pressure have reached the density floor. The intensity of the colour shading signifies the particle saturation in that specific phase space region. After the impact, the mass of the largest remnant is around 13.24 per cent. The square, cross, and diamond symbols mark the state at 1.2, 2.0, and 5.0 h, respectively.

Therefore, the specific impact configuration of the target planet and impactor being the same size and perfectly aligned in their movement direction is not the sole cause of fragmentary disintegration. Fragmentary disintegration is not a phenomenon unique to equal-mass head-on impacts. Instead, it results from a complex interplay of gravitational forces, shock dynamics, thermodynamic processing, and numerical errors.

### 4.5 Effect of the resolution

The results of this study are based on simulations with particle resolution of the target between $10^5$ and $10^6$. As indicated in Meier et al. (2021) and Dou et al. (2024), when resolution exceeds $10^5$, neither the mass nor the iron mass fraction of the largest remnant shows significant dependence on the resolution. For head-on impacts without fragmentary disintegration, we find that the difference in mass and iron mass fraction of the largest remnant is within 3 per cent between particle resolutions of $\sim 10^5$ and $\sim 10^6$. We tested $1.0\,M_\oplus$ equal-mass head-on impacts at target resolutions of $2 \times 10^5$, $5 \times 10^5$, $10^6$, and $10^7$ at $31.06\,\mathrm{km\,s^{-1}}$. Fragmentary disintegration occurred in all tested cases.

### 4.6 Effect of the density floor

The density floor, enforced by a maximum smoothing length $h_{\mathrm{max}}$, is the minimum density that SPH particles can have during simulations, thereby influencing the particles' pressures. A smaller $h_{\mathrm{max}}$ results in a larger density floor (equation 4), causing more particles to reach the density floor, particularly during high-velocity impacts where ejected material moves far apart. When a particle hits the density floor, the density calculation – and consequently the pressure and hydrodynamic force – is affected. This affects the particles' spatial distribution and thermodynamic evolution.

In Fig. 20, the spatial distribution and densities of core particles after a high-velocity impact are shown for various cutoff densities. Particles below 10 times the density floor are coloured in pink. We choose 10 times the density floor to include the particles that could potentially be affected by cut off of the smoothing length and therefore have less neighbours (less than 48 in our simulations). With a small $h_{\mathrm{max}}$, all core particles hit the density cutoff at 10 h (the first panel), while with a large $h_{\mathrm{max}}$, none of the core particles hit the density cutoff (the third and fourth panel). At high-impact velocities, the general post-collision structure is not significantly affected by the density floor set in the SPH simulations. All four post-collision snapshots show void space and central ring-like structures, and fragmentary disintegration occurred at all tested density floors.

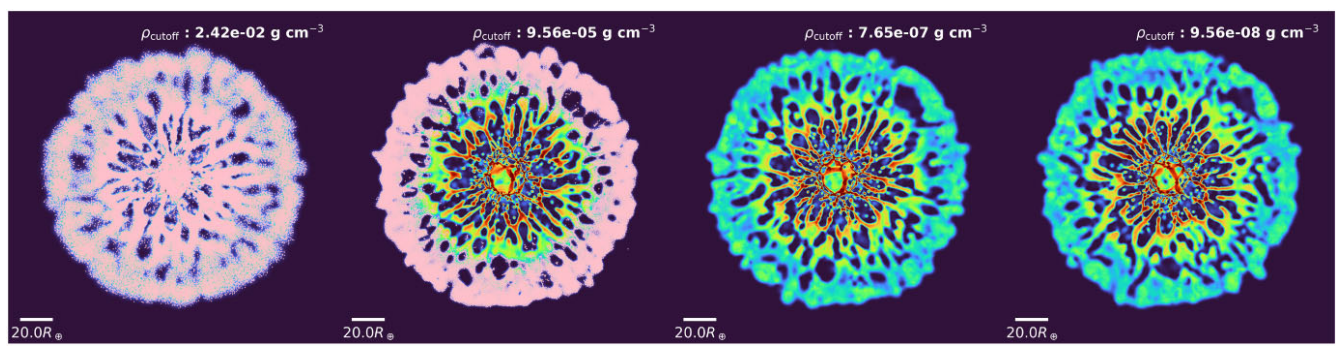

**Figure 20.** The spatial distribution of core material after a 10 h simulation of equal-mass head-on impacts with the target mass of $1.0\,M_\oplus$ at $31.06\,\mathrm{km\,s^{-1}}$. The views are from the impact direction, looking on particles in the $y$–$z$ plane. Coloured representations denote spatial densities, with red indicating denser regions and blue for less dense ones. The pink particles are those that are below 10 times the calculated analytical density floor, as per equation (4). The $\rho_{\mathrm{cutoff}}$ displayed at the top right of each panel represents 10 times the density floor. The larger cutoff density set here is to include particles that might potentially hit the density floor as the simulations progress.

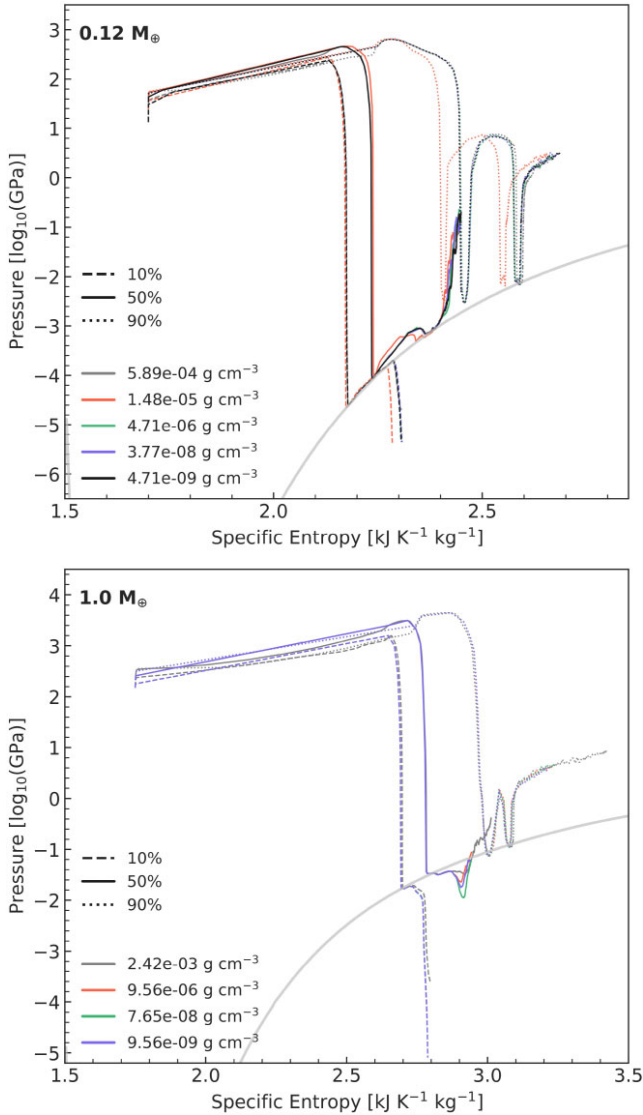

**Figure 21.** The thermodynamic evolution of core material at various density floor during 1–10 h in the entropy and pressure phase diagram for target mass $0.12\,M_\oplus$ (top panel) and $1.0\,M_\oplus$ (bottom panel). The impact velocities are 15.50 and 31.06 km s$^{-1}$, respectively. The three line styles represent the 10, 50, and 90 per cent percentile of the separate tracked entropy and pressure distributions. Each colour shows results from a different density floor. The bold grey line represents partial of the vapour dome of iron EoS we used in the simulations. All the simulations shown here are equal-mass head-on impacts.

In Fig. 21, we demonstrate the variation in the thermodynamic path of core material with different density floors. Density floors generally have minimal influence on the thermodynamic history of core particles, as the thermal paths with different density floors significantly overlap. The lower the density floor, the lower the pressure core particles can reach. However, the location and number of secondary shocks remain similar.

In the top panel of impacts with a target mass of $0.12\,M_\oplus$, the line with a density floor of 1.48e-05 g cm$^{-3}$ (orange-dotted line) is from a simulation with a target resolution of $10^6$, while others are from simulations with a target resolution of $2 \times 10^5$. For the 90 per cent distribution path, the entropies of core particles at high resolution are slightly lower than those of low-resolution simulations, while the pressure scale remains similar. This indicates that resolution could affect the final entropy distribution of head-on impacts, as shocks are resolved at a different scale in higher resolution impact simulations.

In summary, the density floor and resolution of impacts have negligible effects on the thermal history of core particles during head-on impacts. The fragmentary disintegration is not related to the density floor or resolution set in simulations.

## 5 CONCLUSIONS

Head-on impacts are simple but not simple. Head-on impacts simplify the exploration of a vast parameter space while introduce intricate dynamics where shock waves and gravitational forces interact in a complementary manner. Head-on impacts are frequently studied, including in this work, while they are exceedingly rare in the history of planetary formation.

As impact energy increases towards catastrophic levels, head-on collisions exhibit significant variations in the evolution of thermodynamic properties and energy exchange processes. The mass and iron mass fraction of the largest remnant from head-on impacts with different target masses diverge markedly, influenced by varying degrees of shock experienced and the system's gravitational forces. In similar-sized head-on impacts, where the target mass ranges between 0.01 and $1.0\,M_\oplus$, gravitational forces and shock waves interact extensively, causing a substantial number of core particles to interact with the vapour dome. This frequent interaction with the vapour dome and multiple re-shocks could ultimately triggers fragmentary disintegration at lower normalized impact energy than for previously defined super-catastrophic impacts. Our results demonstrate that head-on planetary collisions exhibit distinct behaviours for different target masses, indicating that caution should be exercised when applying scaling laws across a wide parameter space. Head-on impacts may not be the most suitable configuration for investigating the statistical outcomes of giant impacts. Instead, we suggest that more computational resources should be allocated to studying the more prevalent case of oblique impacts, as these collisions are likely to provide a more representative understanding of the consequences of giant impacts in planetary systems.

The exploration of how the varying parameters of impact affect the outcomes continues to be a complex and intriguing field of study. This research further emphasizes the importance of considering the nuances of individual impact scenarios, and not assuming uniformity across different mass and velocity ranges.

In conclusion, our findings shed light on the intricate interplay between shock, gravity, thermodynamics, and core vaporization during high-energy, head-on impacts. The complexity of these interactions, which change significantly across different target masses and impact velocities, underscores the importance of detailed, case-by-case analysis in understanding the outcomes of such catastrophic events. This research provides a stepping stone towards a more nuanced understanding of these processes, which are crucial in modelling and predicting the consequences of large-scale planetary impacts.

ACKNOWLEDGEMENTS

We gratefully acknowledge the reviewer for the insightful comments and constructive suggestions, which significantly improved the quality of this manuscript. JD acknowledges funding support from the Chinese Scholarship Council (No. 202008610218). PJC and ZML acknowledge financial support from the Science and Technology Facilities Council (grant numbers: ST/V000454/1 and ST/Y002024/1). SJL acknowledges support from the UK NERC (grant number: NE/V014129/1). The giant impact simulations were carried out using the computational facilities of the Advanced Computing Research Centre, University of Bristol – http://www.bristol.ac.uk/acrc/ and Isambard 2 UK National Tier-2 HPC Service (http://gw4.ac.uk/isambard/) operated by GW4 and the UK Met Office, and funded by EPSRC (EP/T022078/1).

## DATA AVAILABILITY

Full simulation output is available from the authors on reasonable request.

## APPENDIX A: SIMULATION LIST

**Table A1.** List of simulations involved in Fig. 8.

| $M_{\rm targ}$ (M$_\oplus$) | $V_{\rm imp}$ (km s$^{-1}$) | $M_{\rm lr}/M_{\rm tot}$ | $M_{\rm Fe}/M_{\rm lr}$ | $Q_{\rm R}/Q^*_{\rm RD}$ |
|---|---|---|---|---|
| 0.001 | 2.086 | 0.802 | 0.372 | 0.531 |
| 0.001 | 2.220 | 0.679 | 0.402 | 0.602 |
| 0.001 | 2.353 | 0.635 | 0.412 | 0.676 |
| 0.001 | 2.487 | 0.600 | 0.423 | 0.755 |
| 0.001 | 2.621 | 0.564 | 0.434 | 0.839 |
| 0.001 | 2.755 | 0.532 | 0.444 | 0.927 |
| 0.001 | 2.888 | 0.492 | 0.454 | 1.019 |
| 0.001 | 3.022 | 0.441 | 0.466 | 1.115 |
| 0.001 | 3.156 | 0.388 | 0.480 | 1.216 |
| 0.001 | 3.289 | 0.343 | 0.492 | 1.321 |
| 0.001 | 3.423 | 0.299 | 0.505 | 1.431 |
| 0.001 | 3.557 | 0.271 | 0.514 | 1.545 |
| 0.001 | 3.691 | 0.235 | 0.527 | 1.663 |
| 0.001 | 3.824 | 0.202 | 0.535 | 1.786 |
| 0.001 | 3.958 | 0.174 | 0.544 | 1.913 |
| 0.001 | 4.092 | 0.156 | 0.552 | 2.044 |
| 0.001 | 4.226 | 0.138 | 0.559 | 2.180 |
| 0.001 | 4.359 | 0.126 | 0.568 | 2.321 |
| 0.001 | 4.493 | 0.114 | 0.573 | 2.465 |
| 0.001 | 4.627 | 0.106 | 0.582 | 2.614 |
| 0.001 | 4.760 | 0.096 | 0.582 | 2.767 |
| 0.001 | 4.894 | 0.092 | 0.585 | 2.925 |
| 0.001 | 5.028 | 0.084 | 0.588 | 3.087 |
| 0.010 | 4.508 | 0.816 | 0.366 | 0.523 |
| 0.010 | 4.651 | 0.696 | 0.394 | 0.557 |
| 0.010 | 4.793 | 0.662 | 0.403 | 0.591 |
| 0.010 | 4.935 | 0.650 | 0.406 | 0.627 |
| 0.010 | 5.078 | 0.636 | 0.409 | 0.664 |
| 0.010 | 5.220 | 0.622 | 0.412 | 0.701 |
| 0.010 | 5.362 | 0.604 | 0.417 | 0.740 |
| 0.010 | 5.505 | 0.586 | 0.423 | 0.780 |
| 0.010 | 5.647 | 0.571 | 0.428 | 0.821 |
| 0.010 | 5.790 | 0.553 | 0.435 | 0.863 |
| 0.010 | 5.932 | 0.537 | 0.440 | 0.906 |
| 0.010 | 6.074 | 0.519 | 0.445 | 0.950 |
| 0.010 | 6.217 | 0.501 | 0.451 | 0.995 |
| 0.010 | 6.359 | 0.485 | 0.456 | 1.041 |
| 0.010 | 6.501 | 0.464 | 0.461 | 1.088 |
| 0.010 | 6.644 | 0.441 | 0.468 | 1.136 |
| 0.010 | 6.786 | 0.412 | 0.475 | 1.185 |
| 0.010 | 6.928 | 0.381 | 0.482 | 1.235 |
| 0.010 | 7.071 | 0.342 | 0.491 | 1.287 |
| 0.010 | 7.213 | 0.304 | 0.499 | 1.339 |
| 0.010 | 7.310 | 0.268 | 0.510 | 1.375 |
| 0.010 | 7.406 | 0.236 | 0.520 | 1.412 |
| 0.010 | 7.503 | 0.212 | 0.528 | 1.449 |
| 0.010 | 7.600 | 0.198 | 0.530 | 1.486 |
| 0.010 | 7.696 | 0.176 | 0.538 | 1.524 |
| 0.010 | 7.793 | 0.158 | 0.542 | 1.563 |
| 0.010 | 7.890 | 0.145 | 0.544 | 1.602 |
| 0.010 | 7.986 | 0.138 | 0.541 | 1.641 |
| 0.010 | 8.083 | 0.130 | 0.543 | 1.681 |
| 0.010 | 8.179 | 0.122 | 0.540 | 1.722 |
| 0.010 | 8.372 | 0.039 | 0.517 | 1.804 |
| 0.010 | 8.469 | 0.021 | 0.508 | 1.846 |
| 0.010 | 8.566 | 0.019 | 0.501 | 1.888 |
| 0.120 | 6.434 | 0.978 | 0.306 | 0.206 |
| 0.120 | 7.507 | 0.936 | 0.320 | 0.281 |
| 0.120 | 8.579 | 0.889 | 0.336 | 0.367 |
| 0.120 | 9.115 | 0.867 | 0.345 | 0.414 |
| 0.120 | 9.651 | 0.839 | 0.356 | 0.465 |
| 0.120 | 10.724 | 0.681 | 0.397 | 0.573 |
| 0.120 | 11.260 | 0.648 | 0.406 | 0.632 |
| 0.120 | 11.796 | 0.625 | 0.410 | 0.694 |
| 0.120 | 12.332 | 0.602 | 0.414 | 0.758 |
| 0.120 | 12.868 | 0.571 | 0.419 | 0.826 |
| 0.120 | 13.405 | 0.546 | 0.424 | 0.896 |
| 0.120 | 13.941 | 0.514 | 0.430 | 0.969 |
| 0.120 | 14.477 | 0.480 | 0.438 | 1.045 |
| 0.120 | 15.000 | 0.469 | 0.425 | 1.122 |
| 0.120 | 15.500 | 0.132 | 0.429 | 1.198 |
| 1.000 | 14.421 | 0.977 | 0.306 | 0.267 |
| 1.000 | 16.640 | 0.929 | 0.322 | 0.356 |
| 1.000 | 17.749 | 0.898 | 0.333 | 0.405 |
| 1.000 | 19.967 | 0.824 | 0.362 | 0.512 |
| 1.000 | 20.500 | 0.784 | 0.367 | 0.540 |
| 1.000 | 21.500 | 0.720 | 0.376 | 0.594 |
| 1.000 | 22.186 | 0.696 | 0.383 | 0.632 |
| 1.000 | 23.295 | 0.666 | 0.390 | 0.697 |
| 1.000 | 24.404 | 0.629 | 0.404 | 0.765 |
| 1.000 | 26.623 | 0.546 | 0.425 | 0.910 |
| 1.000 | 27.732 | 0.506 | 0.431 | 0.988 |
| 1.000 | 28.000 | 0.497 | 0.438 | 1.007 |
| 1.000 | 28.842 | 0.454 | 0.456 | 1.069 |
| 1.000 | 29.950 | 0.380 | 0.497 | 1.152 |
| 1.000 | 31.060 | 0.139 | 0.554 | 1.239 |
| 1.000 | 32.170 | 0.043 | 0.620 | 1.329 |
| 11.860 | 31.000 | 0.987 | 0.304 | 0.253 |
| 11.860 | 34.000 | 0.981 | 0.306 | 0.305 |
| 11.860 | 37.000 | 0.965 | 0.311 | 0.361 |
| 11.860 | 40.000 | 0.940 | 0.319 | 0.422 |
| 11.860 | 43.000 | 0.906 | 0.331 | 0.488 |
| 11.860 | 46.000 | 0.861 | 0.349 | 0.558 |
| 11.860 | 50.000 | 0.809 | 0.369 | 0.659 |
| 11.860 | 49.966 | 0.780 | 0.357 | 0.658 |
| 11.860 | 53.000 | 0.728 | 0.377 | 0.741 |
| 11.860 | 55.518 | 0.690 | 0.385 | 0.813 |
| 11.860 | 56.000 | 0.679 | 0.392 | 0.827 |
| 11.860 | 59.000 | 0.624 | 0.414 | 0.918 |
| 11.860 | 61.070 | 0.571 | 0.426 | 0.983 |
| 11.860 | 62.000 | 0.553 | 0.444 | 1.014 |
| 11.860 | 64.000 | 0.506 | 0.465 | 1.080 |
| 11.860 | 65.000 | 0.469 | 0.486 | 1.114 |
| 11.860 | 66.622 | 0.409 | 0.513 | 1.170 |
| 11.860 | 68.000 | 0.376 | 0.544 | 1.219 |
| 11.860 | 71.000 | 0.283 | 0.616 | 1.329 |
| 11.860 | 72.173 | 0.233 | 0.666 | 1.374 |
| 11.860 | 73.337 | 0.186 | 0.721 | 1.418 |
| 11.860 | 74.000 | 0.174 | 0.716 | 1.444 |
| 11.860 | 75.000 | 0.149 | 0.772 | 1.483 |
| 11.860 | 77.000 | 0.048 | 0.860 | 1.563 |

*Note.* All the simulations are equal-mass head-on impacts.

This paper has been typeset from a T$_{\rm E}$X/LAT$_{\rm E}$X file prepared by the author.